\documentclass[twoside,11pt]{article}

\usepackage{jmlr2e} 

\usepackage{bm} 
\usepackage{amsmath}
\usepackage{enumerate} 
\usepackage{dsfont}
\usepackage{booktabs,tabularx,multirow} 
\usepackage{listings}
\usepackage{comment}
\usepackage{color}
\usepackage{subcaption} 
\usepackage{algorithm}
\usepackage{algpseudocode}%
\usepackage{lscape}
\usepackage{afterpage}

\def\E{\mathbb{E}}

\def\Op{O_P}

\DeclareMathOperator{\Cov}{\operatorname{Cov}}
\DeclareMathOperator{\N}{\operatorname{N}}
\DeclareMathOperator{\IG}{\operatorname{IG}}
\DeclareMathOperator{\diag}{\operatorname{diag}}
\DeclareMathOperator{\Logit}{\operatorname{Logit}}
\DeclareMathOperator{\tr}{\operatorname{tr}}

\def\y{\mathbf y}
\def\X{\mathbf X}
\def\x{\mathbf x}
\def\uu{\boldsymbol{\mu}}
\def\rr{\boldsymbol{\gamma}}
\def\bb{\boldsymbol{\beta}}
\def\pp{\boldsymbol{\phi}}
\def\PP{\boldsymbol{\Phi}}
\def\DD{\boldsymbol{\Delta}}
\def\I{\mathbf I}
\def\ee{\boldsymbol{\epsilon}}
\def\b{\mathbf b}
\def\w{\mathbf w}

\def\U{\mathbf U}

\newcommand\numberthis{\addtocounter{equation}{1}\tag{\theequation}}

\makeatletter
\renewenvironment{table}%
  {\selectfont
  \@float{table}}
  {\end@float}
\makeatother

\begin{document}


\title{\bf{A Variational Algorithm for Bayesian Variable Selection}}

\author{\name Xichen Huang \email xhuang43@illinois.edu \\
       \addr Department of Statistics\\
       University of Illinois at Urbana-Champaign\\
       Champaign, IL 61820, USA
       \AND
       \name Jin Wang \email jinwang8@illinois.edu \\
       \addr Department of Statistics\\
       University of Illinois at Urbana-Champaign\\
       Champaign, IL 61820, USA
       \AND
	   \name Feng Liang \email liangf@illinois.edu \\
       \addr Department of Statistics at Urbana-Champaign\\
       University of Illinois\\
       Champaign, IL 61820, USA
	   }

\editor{}

\maketitle 

\begin{abstract}%
\textcolor{blue}{There has been an intense development on the estimation of a sparse regression coefficient vector in statistics, machine learning and related fields.  In this paper, we focus on the Bayesian approach to this problem, where sparsity is incorporated by the so-called spike-and-slab prior on the coefficients. Instead of replying on MCMC for posterior inference, we propose a fast and scalable algorithm based on variational approximation to the posterior distribution. The updating scheme employed by our algorithm is different from the one proposed by \citet{carbonetto12BVS}. Those changes seem crucial for us to show that our  algorithm  can achieve asymptotic consistency even when the feature dimension diverges exponentially fast with the sample size. Empirical results have demonstrated the effectiveness and efficiency of the proposed algorithm.}
\end{abstract}



\section{Introduction}
Consider a standard linear regression problem, where we model $Y$, a continuous response variable, by a linear function of a set of $p$ features $(X_1, \dots, X_p)$ via
\[ Y = X_1 \beta_1 + \cdots X_p \beta_p + \epsilon.   \]
In the past three decades or so, there has been an intense development on the estimation of a sparse regression model. Here ``sparse'' means that only a small fraction of $\beta_j$'s is believed to be non-zero. Identifying the set $S = \{ j: \beta_j \ne 0, j=1, \dots, p\}$ is often referred to as the variable selection problem.

The current approaches to variable selection can be roughly divided into two categories. One category contains approaches based on penalized likelihood, including the classical variable selection procedures like AIC/BIC and the more recent ones like LASSO \citep{Tibshirani94Lasso} and SCAD \citep{Fan01VS}. As the name suggested, the penalized likelihood approach estimates the regression parameter by minimizing the log-likelihood plus some penalty function on $\bb$. With a proper choice of the penalty function, the solution $\hat{\bb}$ will have some of its components to be exactly zero, that is, parameter estimation and variable selection are carried out simultaneously. For an overview of the recent developments on penalized likelihood approaches to variable selection in high dimensions, see \citet{Fan10aselective} and \citet{Buhlmann11shdd}.

We focus on the other category, the Bayesian approach, which starts with a hierarchical prior on all the unknown parameters. 
For example, a widely used prior on $\bb$ is the so-called spike-and-slab prior \citep{Mitchell88BVS}:
\begin{equation} \label{SSprior}
\beta_{j} \vert \gamma_{j}, \sigma_{2} \sim \gamma_{j}\N(0, v_{1}\sigma^2) + (1-\gamma_{j})\delta_{0}, \quad  j=1,\ldots,p,
\end{equation}
where $\delta_{0}$ denotes a point mass at 0, and $\gamma_{j}=1$ if the $j$-th variable is included and $0$ otherwise. The $p$-dimensional binary vector $\rr$, which serves as a model index for all the $2^p$ sub-models,
is then modeled by a product of
 \textit{i.i.d.}\! Bernoulli distributions with parameter $\theta$. 
 
An advantage of the Bayesian approach is that, in addition to the posterior distribution on $\bb$, we can also obtain a posterior distribution on all the sub-models. For example, we can discuss the probability of a sub-model $\rr$ or the inclusion probability of a particular feature, which can be of more interest than a point estimate of $\bb.$  Further, for prediction, it is well-known that model combination or aggregation has a better performance than a single model \citep{Breiman01RF}. The Bayesian approach for variable selection gives rise to a natural averaging scheme: the prediction from various sub-models can be averaged with respect to their posterior probabilities \citep{Raftery98BayesModelAve, Clyde04ModelUncertainty}.

Despite the aforementioned advantages, in practice, Bayesian variable selection is less preferable than those penalization algorithms. A major disadvantage of Bayesian variable selection is the computing cost. The posterior distribution usually does not have a closed-form expression, so posterior inference has to reply on MCMC, which could be time consuming especially when the number of predictors is large.

In this paper, we propose a variational algorithm for Bayesian variable selection. It is a deterministic algorithm, seeking an approximation of the true posterior distribution over $(\bb, \rr)$, instead of running an MCMC chain. It converges very fast and can scale for large sized data sets. Our work is motivated by 
 an earlier variational algorithm proposed by \citet{carbonetto12BVS}. The two algorithms have the  same prior specification and the same set of variational parameters $\Theta$.
The two algorithms, however, update the variational parameters differently. In the algorithm by \citet{carbonetto12BVS}, the parameters associated with each feature are updated sequentially given the others; such a \emph{component-wise updating} scheme is prone to error accumulation especially when $p$ is large and predictors are correlated. In our algorithm, all features are updated simultaneously, which we refer to as the \emph{batch-wise updating} scheme, therefore is more robust to errors and correlations among predictors. Indeed, the batch-wise updating scheme employed by our algorithm turns out to be crucial for us to show our algorithm achieves both frequentist consistency and Bayesian consistency even when $p$ diverges at an exponential rate of the sample size $n$. To the best of our knowledge, no asymptotic results on variational algorithms for Bayesian variable selection are available in the literature.

The remaining of the paper is arranged as follows: Section 2 presents the two variational Bayes (VB) algorithms; Section 3 investigates the asymptotic properties of our new algorithm; Empirical results are given in Section 4 and Section 5, and conclusions are given in Section 6.

\subsection{Notation.}
We define some symbols that will be used in the following sections. For sequences $\{a_{n}\}_{n=1}^{\infty}$ and $\{b_{n}\}_{n=1}^{\infty}$, we write
\begin{itemize}
\item $a_{n} = O(b_{n})$, if $\exists \ c\in \mathds{R}^{+}$ and $n_{0} \in \mathds{N}$, s.t. $| a_{n}/b_{n} | \leq c$, $\forall n \geq n_{0}$;
\item $a_{n} = o(b_{n})$, if $\lim_{n\to\infty} a_{n}/ b_{n}=0$;
\item $a_{n} \asymp b_{n}$, if $\exists \ c_{1},c_{2}\in \mathds{R}^{+}$ and $n_{0} \in \mathds{N}$, s.t. $c_{1}\leq | a_{n}/b_{n} | \leq c_{2}$, $\forall n \geq n_{0}$;
\item $a_{n} \prec  b_{n}$ if $a_{n} = o( b_{n}),$ and $a_{n} \preceq  b_{n}$ if $a_{n} = O( b_{n})$.
\end{itemize}
For a random variable sequence $\{X_{n}\}_{n=1}^{\infty}$ and a constant sequence $\{a_{n}\}_{n=1}^{\infty}$, we write  $X_{n} = \Op(a_{n})$ if $\forall \varepsilon > 0$, $\exists M > 0$ s.t. $P\left(|\frac{X_n}{a_n}| > M \right) < \varepsilon$, $\forall n$, and $X_{n}=o_{p}(a_{n})$ when $\lim_{n \to \infty} P\left(|\frac{X_n}{a_n}| \geq \varepsilon \right) = 0$, $\forall \varepsilon>0$. For $a,b \in \mathds{R}$, we write $a\bigvee b$ to represent the larger number of $a$ and $b$, and $a\bigwedge b$ to represent the smaller one of $a$ and $b$.

\section{Variational Approximation}

\subsection{The Model}
Represent the linear regression model in a matrix form:
\begin{equation} \label{eq:normallinearmodel}
  \y = \X \bb + \ee ,
\end{equation}
where $\ee=(\epsilon_{1},\ldots, \epsilon_{n})^{T}$ is a vector that contains $n$ \textit{i.i.d.}\!  random errors generated from a normal distribution $\N(0, \sigma^{2})$, $\y$ is the response vector of length $n$,  $\X = \left(x_{ij} \right) $ is an $n\times p$ design matrix, and $\bb = (\beta_{1}, \ldots, \beta_{p})^{T}$ is the coefficient vector of length $p$. Like in many other variable selection algorithms,
we center and scale the data as follows:
\[ \sum_i y_i = 0, \quad \sum_i x_{ij} = 0, \quad \sum_i x_{ij}^2 = \|\X_{j}\|_{2}^{2} = n,  \]
where $\X_{j}$ denotes the $j$-th column of $\X$.

The hierarchical prior is specified as follows:
\begin{eqnarray}
 \beta_{j} \vert \gamma_{j} &\sim &\gamma_{j}\N(0, v_{1}\sigma^2) + (1-\gamma_{j})\delta_{0}, \label{eq:normprior:beta}\\
  \gamma_{j} &\overset{i.i.d.}{\sim} & \operatorname{Bern}(\theta), \nonumber  \\
  \sigma^2 &\sim& \IG\left(\frac{\nu}{2}, \frac{\nu\lambda}{2}\right),  \nonumber \\
  \theta &\sim &\operatorname{Beta}(a_{0}, b_{0}),  \nonumber
\end{eqnarray}
where $j=1, \dots, p$, and $\nu$, $\lambda$, $a_0$ and $b_0$ are hyper-parameters.

\subsection{A Variational EM Algorithm}
Variational methods have been widely used in different models, such as the Graphic models \citep{Jordan99VB}. In the ordinary variational Bayesian approach \citep{Bishop06PRML}, an approximating distribution $Q$ of all the latent variables and parameters, which takes a factorized form of $\prod_{j}Q_{j}$, is selected from a restricted family of distributions $\mathcal{Q}$, such that the negative KL-divergence from the true posterior $P$ to $Q$ is maximized, i.e.,
\begin{equation*}
  \max_{Q\in\mathcal{Q}} \E^{Q} \log\frac{P}{Q} = \int\! Q \log\frac{P}{Q}\, \mathrm{d} Q.
\end{equation*}
Then one can solve each $Q_j$ sequentially by fixing other $Q$'s until convergence.

Our variational algorithm is a hybrid of Expectation-Maximization (EM) and variational, same as the one used by \citet{blei2003LDA} for topic models. Next we give a general description of the framework we use for posterior inference.

Let $(\Theta_1, \Theta_2)$ denote the set of parameters of interest, and $\eta$ denote the hyper-parameters. The goal is to obtain an approximation of the posterior distribution on $(\Theta_1, \Theta_2)$. Define the following objective function
\begin{equation}
  \Omega(q_{1}, q_{2}, \eta) = \E_{\Theta_{1},\Theta_{2}}^{q_{1},q_{2}}\log\frac{\pi(\Theta_{1}, \Theta_{2}, \eta \vert  \text{Data})}{q_{1}(\Theta_{1})q_{2}(\Theta_{2})}, \label{eq:KL:vbem}
\end{equation}
where $q_1$ and $q_2$ are distributions on $\Theta_1$ and $\Theta_2$ respectively. Our goal is to find  $q_{1}, q_{2}$, and a point estimate $\hat{\eta}$ to maximize the objective function. We will refer to the estimate $\hat{\eta}$ as the Maximum a posteriori (MAP) estimate: if we optimize (\ref{eq:KL:vbem}) with respect to $q(\Theta_1, \Theta_2)$ and $\eta$, instead of restricting $q(\Theta_1, \Theta_2)$ to take a product form, then the corresponding $\hat{\eta}$ will be exactly the MAP estimate of $\eta.$

Applying the framework above on the Bayesian variable selection model, we estimate the MAP for $\sigma^{2}$ and $\theta$, and approximate the posteriors for $\beta_{j}$'s and $\gamma_{j}$'s. The approximating posterior distribution of $(\bb, \rr)$ takes the following form:
\begin{equation*}
  q(\bb, \rr)=\prod_{j=1}^{p}q_{j}(\beta_{j},\gamma_{j})=\prod_{j=1}^{p}[\phi_{j}f_{j}(\beta_{j})]^{\gamma_{j}} [(1-\phi_{j})\delta_{0}(\beta_{j})]^{1-\gamma_{j}},
\end{equation*}
where $f_{j}(\beta_{j})$ is a probability density function. That is, we approximate the posterior distribution of $\beta_j$ by $\beta_j=0$ with probability $1  - \phi_j$,  and $\beta_j \ne 0$ following a continuous distribution with probability $\phi_j$. 

Given all the information above, we define the following objective function for this problem
\begin{align*}
  \Omega(q_{1}, \ldots, q_{p}, \theta, \sigma^{2})
  &{} = \E^{q_{1}, \ldots, q_{p}}\log \frac{p(\y \vert \bb, \sigma^{2}) p(\bb \vert \rr) p(\rr\vert\theta) \pi(\theta) \pi(\sigma^{2})}
  {\prod_{j=1}^{p}[\phi_{j}f_{j}(\beta_{j})]^{\gamma_{j}} [(1-\phi_{j})\delta_{0}(\beta_{j})]^{1-\gamma_{j}}}.
\end{align*}

\subsection{Algorithm \ref{alg:componetwise} : Component-wise VB}
The first algorithm is similar to the variational algorithm proposed by \citet{carbonetto12BVS}. In detail, we iteratively update the approximating distributions of  $q_{j}(\beta_{j}, \gamma_{j})$'s, and the MAP estimates $\hat{\theta}$ and $\hat{\sigma}^{2}$. Since the algorithm loops over the $p$ dimensions feature by feature, we refer to it as a ``component-wise'' VB algorithm, to highlight its difference with Algorithm \ref{alg:modified},  which we shall propose. 

\subsubsection{Updating Equations}
\begin{description}
\item[Update $q_{j}(\beta_{j},\gamma_{j})$.]
For some $j\in\{1,\ldots,p\}$,  by fixing other approximating distributions and point estimates, we maximize the objective function with respect to $q_{j}$. As shown in \citet{carbonetto12BVS}, 
$f_{j}(\beta_{j})$ is the probability density function of a Normal distribution $\N(\beta_{j} \vert \mu_{j}, \sigma_{j}^{2})$ (albeit we do not assume $f_j$ to be a normal at the beginning) with
    \begin{align*}
      \mu_{j} = & \frac{\X_{j}^{T}\E_{[-j]}\left(\y - \sum_{l\neq j}\X_{l}\beta_{l}\right)}{n+\frac{1}{v_{1}}}, \\
      \sigma_{j}^{2} =& \frac{\hat{\sigma}^{2}}{n+\frac{1}{v_{1}}},
    \end{align*}
    where $\E_{[-j]}$ denotes the expectations over all the $\beta_{l}$'s with $l\neq j$ with respect to the variational distributions.    By symmetry, we know that the other $f_{j}(\beta_{j})$'s are also Normal density functions. As such, we can write $\mu_{j}$ as
    \begin{equation}\label{eq:mu-est-alg1}
      \mu_{j}
      =  \frac{\left(\y - \X_{[-j]}\bar{\bb}_{[-j]}\right)^{T}\X_{j}}{n+\frac{1}{v_{1}}},
    \end{equation}
    where $\X_{[-j]}$ denotes the design matrix without the $j$-th column, and $\bar{\bb} = (\phi_{1}\mu_{1}, \ldots, \phi_{p}\mu_{p})^{T}$ is the mean of $\bb$ w.r.t.\! $q(\bb,\rr)$.

    The log-odds of $\phi_j$  can be updated as
    \begin{equation*}
      \Logit(\phi_{j})=\Logit(\hat{\theta}) + \frac{1}{2}\log\frac{\sigma_{j}^2}{v_{1}\hat{\sigma}^2} +\frac{\mu_{j}^{2}}{2\sigma_{j}^{2}}.
    \end{equation*}

\item[Update  $\hat{\theta}$.] The point estimate of $\theta$ is updated by
    \begin{equation} \label{eq:update:theta}
      \hat{\theta} = \frac{\sum_{j=1}^{p}\phi_{j}+a_{0}-1}{p+a_{0}+b_{0}-2}.
    \end{equation}

\item[Update $\hat{\sigma}^{2}$.]
    The point estimate of $\sigma^{2}$ is updated by
    \begin{equation}
      \hat{\sigma}^{2}
      = \frac{\|\y-\X\bar{\bb}\|_{2}^{2} + \sum_{j=1}^{p}[(n(1-\phi_{j})+1/v_{1})\phi_{j}\mu_{j}^{2} + (n+1/v_{1})\phi_{j}\sigma_{j}^{2}] + \nu\lambda}{n + \prod_{j=1}^{p}\phi_{j} + \nu + 2}. \label{eq:update:sigma2}
    \end{equation}


\end{description}


\begin{algorithm}[htbp]
\begin{algorithmic}
\State \textbf{initialize} $(\mu_{1},\ldots,\mu_{p})$, $(\sigma_{1}^{2},\ldots,\sigma_{p}^{2})$, $(\phi_{1},\ldots,\phi_{p})$, $\hat{\theta}$, and $\hat{\sigma}^{2}$.
\Repeat
    \For{$j$ in $1:p$}
        \State $\mu_{j} \leftarrow \frac{\left(\y - \sum_{l\neq j}\X_{l}\phi_{l}\mu_{l}\right)^{T}\X_{j}}{n+\frac{1}{v_{1}}}$
        \State $\sigma_{j}^{2} \leftarrow \frac{\hat{\sigma}^{2}}{n+\frac{1}{v_{1}}}$
        \State $\phi_{j} \leftarrow \operatorname{Logit}^{-1}\left\{ \log\frac{\hat{\theta}}{1-\hat{\theta}} -\frac{1}{2}\log\frac{v_{1}\hat{\sigma}^{2}}{\sigma_{j}^{2}} +\frac{\mu_{j}^{2}}{2\sigma_{j}^{2}} \right\}$
    \EndFor
    \State $\hat{\theta} \leftarrow \frac{\sum_{j=1}^{p}\phi_{j}+a_{0}-1}{p+a_{0}+b_{0}-2}$
    \State $\hat{\sigma}^{2} \leftarrow \frac{\|\y-\X\bar{\bb}\|_{2}^{2} + \sum_{j=1}^{p}[(n(1-\phi_{j})+1/v_{1})\phi_{j}\mu_{j}^{2} + (n+1/v_{1})\phi_{j}\sigma_{j}^{2}] + \nu\lambda}{n + \prod_{j=1}^{p}\phi_{j} + \nu + 2}$
\Until Converge\\
\Return $(\mu_{1},\ldots,\mu_{p})$, $(\sigma_{1}^{2},\ldots,\sigma_{p}^{2})$, $(\phi_{1},\ldots,\phi_{p})$, $\hat{\theta}$, and $\hat{\sigma}^{2}$
\end{algorithmic}
\caption{Component-wise VB}
\label{alg:componetwise}
\end{algorithm}

\subsubsection{The Drawback of Algorithm \ref{alg:componetwise} in High Dimension} \label{sec:drawback:Alg1}
To reveal the potential drawback of Algorithm \ref{alg:componetwise}  when being applied on a high-dimensional data set, we examine its asymptotic property. 

Assume  the response $\y$ is generated from the normal linear regression model (\ref{eq:normallinearmodel}) with $\bb^{*} \in \mathbb{R}^p$ being the true regression coefficients.  
Consider a relatively easy setting where the minimal eigenvalue of $\X^T \X$ is $O(n)$, i.e., the correlation among columns of $\X$ is small, and our starting values for $\mu_j$'s and $\phi_j$'s are very close to the truth: $\phi_j=0.99$ if $\beta_j^* \ne 0$, $\phi_j=0.01$ if $\beta^*_j = 0$, and
\begin{equation*}
 \mu_j  - \beta^{*}_j =\Op\left(\frac{1}{\sqrt{n}}\right), \text{ for all } j=1, \dots, p.
 \end{equation*}
Then, suppose we are updating the parameters associated with the $j$-th feature $(\mu_j, \sigma^2_j, \phi_j)$. After the update, will $\mu_j$ and $\phi_j$ still be close to the truth?

From Eq \eqref{eq:mu-est-alg1} we have
\begin{align*}
  \mu_{j}
  =& \frac{\X_{j}^{T}(\y-\X_{[-j]}\bar{\bb}_{[-j]})}{n+\frac{1}{v_{1}}}  \\
  =& \frac{nv_{1}}{nv_{1}+1} \Big [  \beta_{j}^{*}  + \frac{1}{n} \X_{j}^{T}\X_{[-j]} \big ( \bb_{[-j]}^{*} - \bar{\bb}_{[-j]} \big ) +  \frac{1}{n}  \X_{j}^T \ee \Big ] .
\end{align*}
Suppose $v_1$ is chosen such that $v_1 n \to \infty$, a condition required for consistency as will be made clear in our analysis in Section \ref{sec:asymtotic}. Then, we have
\begin{equation} \label{mu_j:update}
  \mu_{j} = \beta_{j}^{*} + \Op \left(\frac{p}{\sqrt{n}}\right).
\end{equation}
The result above shows the price we pay for Algorithm \ref{alg:componetwise}: even if we start with $\mu_j$ within a $1/\sqrt{n}$ ball around the truth $\beta_j^*$, after the update, the new $\mu_j$ could be very far away from $\beta_j^*$ when $p$ is large, due to the accumulation of the errors from other dimensions via $\X_{[-j]}\bar{\bb}_{[-j]}$.

Next we examine how $\phi_j$ is affected by the update. Suppose the $j$-th feature is an irrelevant feature, i.e., $\beta_j^*=0.$ The new log-odds of $\phi_j$ is computed as
\begin{align*}
\Logit (\phi_j)
=& \Logit (\hat{\theta}) + \frac{1}{2}\log\frac{\sigma_{j}^2}{v_{1}\hat{\sigma}^2} +\frac{\mu_{j}^{2}}{2\sigma_{j}^{2}} \\
=& O(1)  - \frac{1}{2}\log(v_{1}n+1) + \mu_{j}^{2} \frac{n + \frac{1}{v_{1}}}{2\hat{\sigma}^{2}},
\end{align*}
where $\Logit (\hat{\theta}) = O(1)$ as long as we do not start with $\hat{\theta}=0$ or $1.$ Since $\mu_{j}^{2} =  \Op\left (p^2/n \right)$ by (\ref{mu_j:update}), we have
\begin{equation} \label{eq:logit:noise:asymp:alg1}
2  \Logit( \phi_{j} )  =  - \log(v_{1}n+1) + \Op(p^2). 
\end{equation}
When $p$ is very large, the right hand side of (\ref{eq:logit:noise:asymp:alg1}) may be positive. That is, the  new $\phi_j$ could be bigger than $0.5$, although we start with $\phi_j=0.01$, a value that is very close to the truth. 

Our analysis above is not rigorous, but it clearly reveals an issue with Algorithm \ref{alg:componetwise}: the noise can accumulate due to the feature by feature updating scheme.
 To address this issue, we propose another algorithm which updates $(\mu_1, \dots, \mu_p)$ simultaneously for all $p$ features. 

\subsection{Algorithm \ref{alg:modified}: Batch-wise VB}
Recall the variational parameters we need to update are $\{\mu_j, \phi_j, \sigma^2_j \}_{j=1}^p$. At iteration $t$, instead of updating the triplet $\{\mu_j, \phi_j, \sigma^2_j \}$ sequentially for each $j$ as in Algorithm \ref{alg:componetwise}, we consider the following batch-wise update: update $\{\sigma_{j}^{2}\}_{j=1}^{p}$, then update $\{\mu_{j}\}_{j=1}^{p}$, and finally update $\{\phi_{j}\}_{j=1}^{p}$.
Given $\{\mu_j, \phi_j, \sigma^2_j \}_{j=1}^p$, we can update $(\hat{\theta}, \hat{\sigma}^2)$ using Eq (\ref{eq:update:theta}) and Eq (\ref{eq:update:sigma2}).

\subsubsection{Updating Equations}
\begin{description}
  \item[Update $\{\sigma_{j}^{2}\}_{j=1}^{p}$.] We update  $\sigma_{j}^{2}$'s 
  by maximizing
   \begin{eqnarray*}
      &  &  \E^{q_{1}, \ldots, q_{p}} \left\{
      \sum_{i=1}^{n}\left[-\frac{(y_{i} - \x_{i}^{T}\bb)^2}{2\sigma^{2}} \right]
      + \sum_{j=1}^{p} \gamma_{j} \left[-\frac{\beta_{j}^{2}}{2v_{1}\sigma^{2}} - \log[f_{j}(\beta_{j})]\right]
      \right\} \\
      & \propto & \sum_{j=1}^{n} -\phi_{j}\frac{n}{2\hat{\sigma}^{2}}\sigma_{j}^{2} - \phi_{j}\frac{1}{2v_{1}\hat{\sigma}^{2}} \sigma_{j}^{2} + \frac{1}{2}\phi_{j}\log(\sigma_{j}^{2}),
      \end{eqnarray*}
  and the updating equation for $\sigma_j^2$ is
  \begin{equation} \label{eq:alg2-sigmaj-old}
   \sigma_{j}^{2} = \frac{\hat{\sigma}^{2}}{n + \frac{1}{v_{1}}}, \quad j=1,\ldots,p.
\end{equation}
    Note that $\sigma_{j}^{2}$'s take the same form for all $j$. The term $n$ in the denominator is due to the fact that each column of $\X$ has been pre-processed such that $\|\X_{j}\|^{2}=n$. Later in Section \ref{sec:asymtotic}, in light of the asymptotic analysis, we will suggest to replaced $n$ by $a_n \asymp n^{a}$ as in Eq (\ref{eq:alg2-sigmaj-new}). 

  \item[Update $\{\mu_{j}\}_{j=1}^{p}$.]
    We update $\mu_{j}$'s 
    by maximizing
   \begin{equation*}
    \E^{q_{1}, \ldots, q_{p}} \left\{
      \sum_{i=1}^{n}\left[-\frac{(y_{i} - \x_{i}^{T}\bb)^2}{2\sigma^{2}} \right]
      + \sum_{j=1}^{p} \gamma_{j} \left[-\frac{\beta_{j}^{2}}{2v_{1}\sigma^{2}} - \log[f_{j}(\beta_{j})]\right]
      \right\},
      \end{equation*}
     and the updating equation for $\uu$ is
  \begin{equation} \label{eq:mu-est-alg2}
  \uu = \left(\PP\X^{T}\X\PP + \DD + \frac{1}{v_{1}} \PP \right)^{-1} \PP\X^{T}\y,
    \end{equation}
    where $\DD = \diag\{\X^{T}\X\}\PP(\I-\PP) = n\PP(\I-\PP)$.

  \item[Update $\{ \phi_{j}\}_{j=1}^p$.] The objective function at this step involves a quadratic form of $\pp = (\phi_{1},\ldots, \phi_{p})^{T}$. To simplify the computation, we apply a linear approximation to replace the quadratic term. The detailed derivation can be found in Appendix A, and
  the final updating equation for $\phi_j$ is given by
\begin{align*}\label{eq:phi-est-alg2}
\Logit(\phi_{j}) =& \Logit(\hat{\theta}) + \frac{1}{2}\log\frac{\sigma_{j}^{2}}{v_{1}\hat{\sigma}^{2}} + \frac{\mu_{j}^{2}}{2\hat{\sigma}^{2}}\left(n+\frac{1}{v_{1}}\right) \\
=& \Logit(\hat{\theta}) + \frac{1}{2}\log\frac{\sigma_{j}^{2}}{v_{1}\hat{\sigma}^{2}} + \frac{\mu_{j}^{2}}{2\sigma_{j}^{2}}. \numberthis
\end{align*}

  \item[Truncate $\phi_{j}$'s.]  The final expression of $\phi_j$ involves the exponential of $\Logit(\phi_{j})$, which could trigger the error of numerical overflow when the magnitude of the Logit value is large. In our implementation, we truncate the logit value, or equivalently restrict $\phi_{j}^{(t)} \in [c, 1-c],$ where  $1>c>0$ is a small constant. 
  
  We also stop updating any $\phi_{j}$'s once they reach the extreme values, $c$ or $1-c$. That is, for $t>1$
    \begin{equation} \label{eq:alg2-phi-update}
      \phi_{j}^{(t)} =
      \left\{
      \begin{array}{lc}
        \operatorname{Logit}^{-1} \left(\log\frac{\hat{\theta}}{1-\hat{\theta}} + \frac{1}{2}\log\frac{\sigma_{j}^2}{v_{1}\hat{\sigma}^2} +\frac{\mu_{j}^{2}}{2\sigma_{j}^{2}}\right), & \mbox{if } \min\left(\phi_{j}^{(t-1)}, 1-\phi_{j}^{(t-1)}\right) > c; \\
        \phi_{j}^{(t-1)}, &  \mbox{if } \min\left(\phi_{j}^{(t-1)}, 1-\phi_{j}^{(t-1)}\right) \leq c.
      \end{array}
      \right.
    \end{equation}
This stop-early updating scheme can dramatically reduce the computation cost for our algorithm, as explained in Section \ref{sec:comp:complexity}. 

\item[Stopping Criterion.]
After we loop over all the parameters mentioned above, we need to decide when to stop. A natural choice is to stop when the change of the objective function is less than some threshold. Since our primary focus is variable selection, we use the maximum entropy criterion: we compute the entropy for each $\text{Bern}(\phi_j)$, and stop if the maximum of the change of the entropy is less than a pre-specified threshold.
\end{description}

\begin{algorithm}[htbp]
\caption{Batch-wise VB}
\label{alg:modified}
\begin{algorithmic}
\State \textbf{Initialize} $(\mu_{1},\ldots,\mu_{p})$, $(\sigma_{1}^{2},\ldots,\sigma_{p}^{2})$, $\hat{\theta}$, and $\hat{\sigma}^{2}$.
\State \textbf{Initialize} $(\phi_{1},\ldots,\phi_{p}) = \mathbf{1}_{p\times 1}$, and truncation parameter $c$.
\Repeat
    \State $\uu \leftarrow (\PP\X^{T}\X\PP + \DD + \frac{1}{v_{1}} \PP)^{-1} \PP\X^{T}\y$
    \State $\sigma_{j}^{2} \leftarrow \frac{\hat{\sigma}^{2}}{a_n+1/v_{1}} \text{ or } \frac{\hat{\sigma}^{2}}{n+1/v_{1}}$
    \For{$j$ in $1:p$}
        \If {$\min\left(\phi_{j}, 1-\phi_{j}\right) > c$}
            \State $\phi_{j} \leftarrow \operatorname{Logit}^{-1}\left\{ \Logit(\hat{\theta}) + \frac{1}{2}\log\frac{\sigma_{j}^{2}}{v_{1}\hat{\sigma}^{2}} + \frac{\mu_{j}^{2}}{2\sigma_{j}^{2}} \right\}$
        \EndIf
    \EndFor
    \State $\hat{\theta} \leftarrow \frac{\sum_{j=1}^{p}\phi_{j}+a_{0}-1}{p+a_{0}+b_{0}-2}$
    \State $\hat{\sigma}^{2} \leftarrow \frac{ \| \y-\X\PP\uu \|^2 + \sum_{j=1}^{p}[(\X_{j}^{T}\X_{j}(1-\phi_{j})+1/v_{1})\phi_{j}\mu_{j}^{2} + (\X_{j}^{T}\X_{j}+1/v_{1})\phi_{j}\sigma_{j}^{2}] + \frac{1}{v_{1}}\sum_{j=1}^{p}\phi_{j}(\mu_{j}^{2} + \sigma_{j}^{2}) + \nu\lambda}{n + \prod_{j=1}^{p}\phi_{j} + \nu + 2}$
\Until Converge\\
\Return $(\mu_{1},\ldots,\mu_{p})$, $(\sigma_{1}^{2},\ldots,\sigma_{p}^{2})$, $(\phi_{1},\ldots,\phi_{p})$, $\hat{\theta}$, and $\hat{\sigma}^{2}$
\end{algorithmic}
\end{algorithm}

\subsubsection{Computational Complexity} \label{sec:comp:complexity}
The main computational cost lies in reverting a $p\times p$ matrix in Eq \eqref{eq:mu-est-alg2}. The direct computation  involves $O(p^3\bigwedge n^3)$ operations which is time-consuming when both $p$ and $n$ are large. Next we describe the computation trick used in our implementation, which can dramatically reduce the computation cost.

At iteration $t$,  Eq \eqref{eq:mu-est-alg2}  can be rewritten as
\begin{eqnarray*}
\uu &=&  \left(\X^{T}\X\PP^{(t)} + n(\I - \PP^{(t)}) + \frac{1}{v_{1}}\I \right)^{-1} \X^{T}\y \\
&=&  \left(A_{(t-1)} + \big (A_{(t)} - A_{(t-1)} \big ) \right)^{-1} \X^{T}\y,
\end{eqnarray*}
where $A_{(t)} = \Big (\X^{T}\X\PP^{(t)} + n(\I - \PP^{(t)}) + \frac{1}{v_{1}}\I \Big)$ and
\[
A_{(t)} - A_{(t-1)} =   \big (\X^{T}\X- n\I_{p} \big ) \big (\PP^{(t)}-\PP^{(t-1)} \big).
\]
At iteration $t>1$, we would have $A_{(t-1)}^{-1}$ in hand. If the rank of $A_{(t)} - A_{(t-1)}$ is lower than $p$ or $n$, then the problem can be reformulated as inverting a matrix  of a lower rank. 

Write $\X^{T}\X- n\I_{p} = B$ and $D^{(t)} = \PP^{(t)}-\PP^{(t-1)}$. Then $A_{t} - A_{(t-1)} = BD^{(t)}$. 
Based our experience, after the first several iterations, many $\phi_j$'s are close to $1$ or $0$, i.e.,  they will not be updated any more according to Eq \eqref{eq:alg2-phi-update}. So many diagonal elements of $D^{(t)}$ are zero. Without loss of generality, assume only the first $q$ elements of $D^{(t)}$ are not zero. Then we have $BD^{(t)}= UCV$, where $U = B_{[,1:q]}$ contains only the first $q$ columns of $B$, $C = \diag\{\phi_{j}^{(t)}-\phi_{j}^{(t-1)}\}_{j=1}^{q}$, and $V=(\I_{q}, \mathbf{0}_{q\times(p-q)})$. Applying the Woodbury formula, we have
\[
\uu = \left(A_{(t-1)}^{-1} - A_{(t-1)}^{-1}U \left(C^{-1}+VA_{(t-1)}^{-1}U \right)^{-1}_{q \times q} VA_{(t-1)}^{-1} \right) \X^{T}\y.
\]
So  we only need to invert a $q \times q$ matrix where  $q$ is much smaller than $p$ and $n$.

\section{Asymptotic Analysis}  \label{sec:asymtotic}

Assume  the response $\y$ is generated from the normal linear regression model (\ref{eq:normallinearmodel}) with $\bb^{*} \in \mathbb{R}^p$ being the true regression coefficients.  Let $\rr^*$ denote the true model index, i.e., $\gamma^*_j = 1$ if $\beta_j^* \ne 0$ and $\gamma^*_j = 0$ if $\beta_j^* = 0$. Also define the true set of relevant variables as 
$$S^* = \{j: \beta^*_j \ne 0\} = \{j: \gamma^*_j = 1\}. $$
In our analysis, the dimension $p=p_n$ is allowed to diverge with $n$, and therefore $\bb^*=\bb^*_n$, $\rr^* = \rr^*_n$ and $S^* = S^*_n$ may also vary with $n$.


We will show that Algorithm \ref{alg:modified} achieves both the frequentist  consistency and the Bayesian consistency. Recall that Algorithm \ref{alg:modified} returns an estimate of the relevant variable set via
\[ \hat{S}_n = \{j: \phi_j > 0.5\}. \]
The frequentist consistency refers to the convergence (in probability) of $\hat{S}_n$ to $S^*$.
The cut-off value $0.5$ can be changed to any other fixed value in $(0,1)$, which will not affect the consistency result as shown in our analysis.

In addition to a point estimate of the true variable set,  our algorithm also returns a probability distribution over all $2^p$ variable sets (or sub-models), namely,
\[ q(\rr) =  \prod_{j=1}^p \big ( \phi_{j} \big)^{\gamma_j} \big ( 1 - \phi_{j} \big )^{1-\gamma_j}. \]
The aforementioned frequentist consistency corresponds to $q(\rr^*)$ is the largest, i.e., the truth model receives the largest posterior probability, while the Bayesian consistency requires $q(\rr^*)$ converges to $1$ in probability, which is stronger than the frequentist consistency.

In addition to the ordinary regularity conditions, for our algorithm to achieve consistency, we need to let $v_1$, the prior variance on the non-zero $\beta_j$'s as in Eq (\ref{eq:normprior:beta}), to grow to infinity at a certain rate of $n$. A similar condition also arises in the asymptotic study on Bayesian variable selection by \citet{Naveen:He:2014} although their prior specification is different from ours. 
To help the readers to understand this condition, we first give the asymptotic analysis on a simple orthogonal design and then describe the general result.

\subsection{The Orthogonal Design}
Consider a simple case in which the design matrix is orthogonal, i.e., $\X^{T}\X = n$. To simplify our discussion, we also assume that $\sigma^2$ is known, $\theta$ is set to be $1/2$, and the minimal non-zero coefficient is bigger than some constant (i.e., the non-zero coefficients will not diminish to zero). These conditions will be relaxed in our result for the general case.

Suppose we run our algorithm for one step. From the updating equations of Algorithm \ref{alg:modified}, we have
\begin{equation} \label{Logit:orthogonal}
2\Logit(\phi_{j}) = -\log(v_{1}n+1) + \frac{n\hat{\beta}_{j}^{2}}{\sigma^{2}}\frac{n}{n+\frac{1}{v_{1}}},
\end{equation}
where $\hat{\beta}_{j}=\frac{1}{n}\X_{j}^{T}\y \sim\N\Big (\beta_{j}^{*}, \frac{\sigma^{2}}{n}\Big)$
is the OLS estimator of the $j$-th coefficient. 

When  $p$ is fixed, as in the classical asymptotic setting, it is easy to show  that our algorithm has the desired asymptotic behavior as long as
\begin{equation} \label{cond:orthogonal:fixed_p}
v_1 n \to \infty, \quad  \log( v_1 n) = o(n).
\end{equation}
This is because:  when $\beta_{j}^{*}=0$, since $n \hat{\beta}_{j}^{2} = \Op(1)$, the leading term in (\ref{Logit:orthogonal}) is the first term that goes to $-\infty$, therefore $\phi_j$ goes to $0$; when $\beta_{j}^{*} \ne 0$,  the leading term in (\ref{Logit:orthogonal}) is the second term that goes to $\infty$, therefore $\phi_j$ goes to $1$. 

When $p=p_n$ increases with $n$,  the coefficients $(\beta_{j}^{*})_{j=1}^{p}$ and  the true variable set $S_n^*$ may vary with $n$.
As such, it is no longer meaningful to discuss the limit of Eq (\ref{Logit:orthogonal}).
Instead, we need to examine the limiting behavior of $\max_{j\notin S_{n}^{*}}\Logit(\phi_{j}) $ and $\min_{j\in S_{n}^{*}}\Logit(\phi_{j}) $. 


First we show that the frequentist variable selection consistency could be achieved with $p = O(\sqrt{v_1 n})$, in addition to condition (\ref{cond:orthogonal:fixed_p}). Let $C$ be an arbitrary positive number.  It suffices to show that
\begin{equation} \label{eq:prob-sum=0}
 P\left( \max_{j\notin S_{n}^{*}}\Logit(\phi_{j}) > -\frac{C}{2}  \right) + P\left( \min_{j\in S_{n}^{*}}\Logit(\phi_{j}) < \frac{C}{2} \right) \to 0.
 \end{equation}
By the Bonferroni correction and the tail probability inequality of the normal distribution, we have
\begin{eqnarray}
& & P\left( \max_{j\notin S_{n}^{*}}\Logit(\phi_{j}) > - \frac{C}{2} \right) \nonumber \\
&\leq & \sum_{j\notin S_{n}^{*}} P\left(\frac{1}{\sigma^{2}}n\hat{\beta}_{j}^{2}\frac{n}{n+\frac{1}{v_{1}}} >  \log(v_{1}n+1) - C \right) \nonumber \\
& \leq &  \frac{c}{\sqrt{\log(v_{1}n) - C}} \exp\left\{-\frac{\log(v_1 n) - C}{2} + \log p  \right\} \to 0, \label{eq:nonsignal:tailprob}
\end{eqnarray}
as long as $p =O(\sqrt{v_{1}n})$ and $v_{1}n \to \infty$,
where $c$ is some constant.
Similarly, we have
\begin{align*}
& P\left( \min_{j\in S_{n}^{*}}\Logit(\phi_{j}) < \frac{C}{2} \right)
\leq  P\left(\frac{1}{2}\min_{j\in S_{n}^{*}}\frac{1}{\sigma^{2}}n\hat{\beta}_{j}^{2} < \log(v_{1}n+1) + C \right)\\
\leq & P\left(\frac{1}{\sigma}\sqrt{n} \min_{j\in S_{n}^{*}}\vert\beta_{j}^{*}\vert - \frac{1}{\sigma}\sqrt{n} \max_{j\in S_{n}^{*}}\vert\hat{\beta}_{j}-\beta_{j}^{*}\vert < \sqrt{2\log(v_{1}n+1)+ 2 C}\right)\\
\leq &
 | S_{n}^* | P\left(\frac{1}{\sigma}\sqrt{n}\vert\hat{\beta}_{j}-\beta_{j}^{*}\vert > r \sqrt{n}
 \right),
\end{align*}
which also goes to zero by the tail probability of the normal distribution, where $r$ is some constant.

Secondly, we show that the Bayesian consistency could be achieved with $p = o(\sqrt{v_1 n})$.
We can show that, if $p=o(\sqrt{v_1 n})$, Eq (\ref{eq:prob-sum=0}) still holds with a varying constant $C_{n}=s\log(v_{1}n)$ where $s \in (0,1)$. 
Then with probability going to 1, we have
\[ \max_{j\in S_{n}^{*}}(1-\phi_{j}) \bigvee \max_{j\notin S_{n}^{*}} \phi_{j} < \frac{1}{\exp\{\frac{1}{2}C_{n}\}} < \frac{1}{(v_{1}n)^{s/2}}. \]
Using the inequality $\prod_{j}(1-p_{j}) \geq 1-\sum_{j} p_{j}$, we have
\begin{align*}
q_(\rr^*) & = \Big [ \prod_{j: \gamma^*_{j} =1} \phi_{j} \Big] \Big [ \prod_{j: \gamma^*_{j} =0} (1 - \phi_{j}) \Big ]
\geq  1-  \sum_{j\in S_{n}^{*}}(1-\phi_{j}) + \sum_{j\notin S_{n}^{*}} \phi_{j} \\
\geq &  1 - \frac{p}{(v_{1}n)^{s/2}} \to 1.
\end{align*}

Our analysis for the orthogonal case indicates that the choice of $v_{1}$ affects how fast we could allow $p$ to diverge with $n$. For example, if $v_{1}$ is a constant, then our algorithm can achieve the frequentist consistency for $p = O(\sqrt{n})$ and Bayesian consistency for $p = o(\sqrt{n})$. In order to achieve consistency for larger $p$,  we need to let $v_{1}\to \infty$ with $n$.

\subsection{The General Case}

Without loss of generality, assume $S_n^*=\{1, ..., q_n \}$ and the true coefficient $\bb^{*} = (\bb^{*T}_{1}, 0^T)^T$, i.e, the first $q_n$ features are the relevant ones. Write the design matrix as $\X=(\X_1,\X_2)$ accordingly, where $\X_1$ is the $n\times q_n$ matrix corresponding to the signal features and $\X_2$ is the $n \times (p-q_n)$ matrix corresponding to the noise features.

In our analysis, we assume the following conditions hold. 
\begin{enumerate}[(C1)]
\item  Condition on model identification: Denote $H_1$ and $H_2$ as the projection matrices of $\X_1$ and $\X_2$ onto their column spaces respectively, then assume the rank of $H_1$ is $q_n$ and the spectral norm of $H_1 H_2$ is upper bounded by $1$, i.e., $ \| H_1H_2 \|_2<1.$

\item  Condition on the design matrix: Let $\lambda_{n1}$ denote the minimal non-zero eigenvalue of matrix $\X^t \X$ and it satisfies
\[ \lambda_{n1}^{-1} = O(n^{ -\eta_1}), \quad  0 < \eta_1 \le 1.  \]

\item Condition on the sparsity of $\bb^{*}$: The $L_2$ norm of the true regression coefficient $\bb^*$ satisfies the following sparsity condition
\[ \| \bb^* \|_{2}^{2} = O(n^{\eta_{2}}), \quad  0 \leq \eta_{2} < \eta_{1}. \]

\item The beta-min condition: The  minimal non-zero coefficient  satisfies
\begin{equation}\label{eq:beta*-limit}
\liminf_n \frac{  \min_{j \in S^*_n} \sqrt{n}| \beta^*_{j} | }{n^{\eta_{3}/2}} \ge M,
\end{equation}
where $\eta_{3} \in (1-\eta_{1}, 1]$ and $M > 0$ are two constants not depending on $n$.

\item Condition on the initial values: the initial value for all the inclusion probability $\phi_{j}^{(0)}$'s
should be set to be 1, i.e., we start our algorithm with \textit{all the variables in}. The initial values for the error variance $\hat{\sigma}^{2}$ and the Bernoulli parameter $\hat{\theta}$ could be any constants satisfying $0 < \hat{\sigma}^{2} < \infty$ and $0 < \hat{\theta} < 1$. For the proof, we set
\begin{equation}
\phi_{j}^{(0)} = 1, \quad \hat{\sigma}^{2(0)}=1, \quad  \hat{\theta}^{(0)} = \frac{1}{2}.
\end{equation}
\end{enumerate}

In general, it is not realistic to derive consistent variable selection procedures and  parameter estimation  when the design matrix $\X$  is not of full rank \citep{Shao12}. This is because the true coefficient $\bb^*$ could be any vector from the set  $\mathcal{B} = \{ \bb: \X \bb = \X \bb^* \}$ due to the collinearity among the columns of $\X$. Condition (C1) ensures that the true sparse coefficient $\bb^*$ is identifiable. Let $\bb=(\bb_1^T,\bb_2^T)^T$ be any vector from $\mathcal{B}$. Then we have $\X_1 \bb_1^* = \X_1 \bb_1 + \X_2 \bb_2$. Meanwhile (C1) implies that $\X_1  \bb_1^* = \X_1 \bb_1$. So $\| \bb \|_2 \ge \| \bb^*\|_2$ with equality holds true only if $\bb  = \bb^*$. That is, the true coefficient vector $\bb^*$ is the one from $\mathcal{B}$ with the smallest $L_2$ norm, which is unique.

\begin{figure}[htbp]
  \centering
  \includegraphics[width=0.4\textwidth]{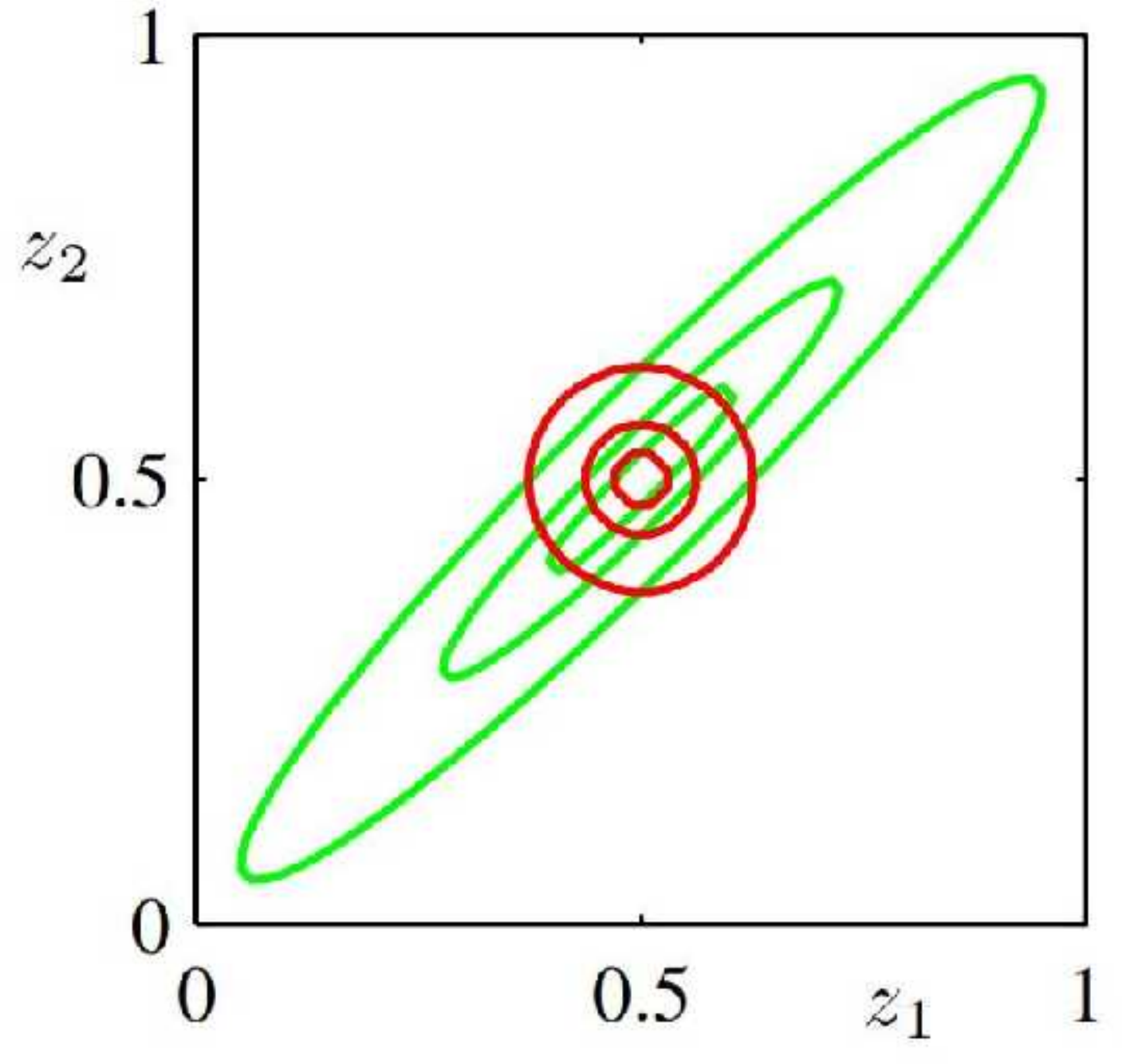}\\
  \caption{Page 468 from \citet{Bishop06PRML}: The green contours correspondens to the 1, 2, and 3 standard deviations for a correlated two-dimensional Gaussian distribution $p(z_1, z_2)$, and the red contours represent the corresponding levels for an approximating distribution $q_1(z_1)q_2(z_2)$ where $q_1$ and $q_2$  are obtained by minimization of the Kullback-Leibler  divergence $KL(q\|p)$.}\label{fig:VB-Bishop}
\end{figure}

In our algorithm, we approximate the posterior distribution of $\beta_j$ by a mixture of a point mass at zero and a normal distribution. The updating equation \eqref{eq:alg2-sigmaj-old} implies that the posterior variance $\sigma_j^2$ for non-zero $\beta_j$'s is of order $1/n$, a reasonable result in the classical asymptotic setting where $p$ is fixed and the minimal eigenvalue of $\X^T \X$ is of $O(n)$. However, in the diverging $p$ case, as indicated by (C2), the minimal non-zero eigenvalue of $\X^T \X$ could be of order $O(n^{\eta_1})$. Therefore the posterior variance $\sigma^2_j$ defined in Eq \eqref{eq:alg2-sigmaj-old} would be too small. In other words, we are too optimistic about the uncertainty of $\beta_j$. This is a common issue with variational  algorithms, as pointed out in \cite{Bishop06PRML} and shown in Figure \ref{fig:VB-Bishop}: the variational distribution  tends to have a smaller support than the true target distribution. To fix this problem, we need to correct the posterior variance at the right order as follows:
\begin{equation}\label{eq:alg2-sigmaj-new}
    \sigma_{j}^{2}=\frac{\hat{\sigma}^{2}}{a_n+\frac{1}{v_{1}}},
\end{equation}
where $a_n \asymp n^{a}$ with $1- \eta_3 < a < \eta_1. $

Below we summarize conditions on various rate parameters appearing in the our assumptions:
\begin{eqnarray}
0 & \le & \eta_2 < \eta_1  \le 1,  \label{eq:rate:cond1} \\
 0 & \le  & 1 - \eta_3 < a < \eta_1 \le 1, \label{eq:rate:cond2}
\end{eqnarray}
where (\ref{eq:rate:cond1}) also appears in other work on variable selection, such as \citet{Shao12}, and (\ref{eq:rate:cond2}) indicates that  the magnitude of the posterior variance (of order of $1/n^{a}$) should be between the minimal signal (of order $1/n^{1-\eta_3}$) and the maximal noise level (of order $1/n^{\eta_1}$). 
In the classical asymptotic setting, we have $\eta_1 = 1, \eta_2=0$ and $1 - \eta_3=0.$

With the modified $\sigma_{j}^{2}$ and some proper choice of $v_{1}$, we can show that our algorithm achieves the desired asymptotic property. We first present a lemma that shows that
 when the sample size is large enough, after one step of Algorithm \ref{alg:modified}, there is a gap between $\max_{j \notin S^*_n}  \Logit(\phi_{j})$ and $\min_{j \in S^*_n}  \Logit(\phi_{j})$ which is large enough for us to separate the relevant variables from the irrelevant ones. Using this lemma, we can then prove the frequentist consistency and the Bayesian consistency. We present our asymptotic results below and include the proofs  in Appendix B. 

\begin{lemma} \label{lemma:one-step-gap}
Assume conditions (C1-C5). Suppose $v_1$ is chosen such that
$$ \Big ( n^{-a} \bigvee n^{- (2 \eta_1 -a- \eta_2)/2} \Big) \prec v_1 \prec e^{n^{(a + \eta_3 -1)}}$$ and $p=p_n$ satisfies $\log p_n = o(n^{\eta_1 - a})$, then for any constant $C >0$, after one step of Algorithm \ref{alg:modified}, we have
\begin{align}
P\left( \max_{j\notin S_{n}^{*}}\Logit(\phi_{j}) > - \frac{C}{2}\right) \longrightarrow 0, \label{lemma:prob:noise} \\
P\left( \min_{j\in S_{n}^{*}}\Logit(\phi_{j}) < \frac{C}{2}\right) \longrightarrow 0. \label{lemma:prob:signal}
\end{align}
\end{lemma}

\begin{theorem}(Variable Selection Consistency) \label{thm:one-step-frequentist}
Assume all conditions in Lemma \ref{lemma:one-step-gap}, then we have
\[
P(\hat{S}_{n} = S_{n}^{*}) \longrightarrow 1.
\]
\end{theorem}

\begin{theorem}(Bayesian Consistency) \label{thm:one-step-bayesian}
Assume conditions (C1-C5). Suppose $v_1$ is chosen such that
$$ \frac{p^2}{a_n} \prec v_1 \prec e^{n^{(a + \eta_3- 1)}}$$ and $p=p_n$ satisfies $\log(p_n) = o(n^{\frac{a + \eta_3-1}{2}} \bigwedge n^{\eta_1 - a})$, then we  have
\[q(\rr^*) = \left[ \prod_{j: \gamma^*_{j} =1} \phi_{j} \right] \left[ \prod_{j: \gamma^*_{j} =0} (1 - \phi_{j}) \right] \overset{P}{\longrightarrow} 1.\]
\end{theorem}


\section{Simulation Studies}
In this section, we conduct several simulation studies to compare the two VB algorithms: Algorithm \ref{alg:modified} with Algorithm \ref{alg:componetwise}, and some other commonly used methods, like LASSO and SCAD. 

In both VB algorithms, the choice of $v_1$ is chosen by cross-validation, and a sparse estimate of $\bb$ is given by 
\begin{equation} \label{eq:sparse:beta}
\hat{\beta}_j = \mu_j \text{ if } \phi_j \ge 0.5; \quad 0 \text{ if } \phi_j < 0.5.
\end{equation} 
The value of $a_n$ in the new variance update formula  (\ref{eq:alg2-sigmaj-new}) is set to be $\lambda_{n1}$, the minimal non-zero eigenvalue of the sample covariance matrix. In our asymptotic analysis we set $a_n$ to be of a smaller order of $\lambda_{n1} $; the main purpose of this choice is to ensure that the dimension $p_n$ can grow exponentially fast,  i.e., $\log p_n = o(n^{\eta_1 - a})$. In practice we have found that setting $a_n$ to be $\lambda_{n1}$ work well, along with the adaptive choice if $v_1$ via cross-validation.

\subsection{Example 1: Benchmark Data}
This is a popular benchmark data set, initially designed by \citet{Tibshirani94Lasso} and latter used by \citet{Fan01VS} to compare different variable selection methods. The true coefficient  is $\bb^* = (3,1.5,0,0,2,0,0,0)^{T} \in \mathbb{R}^8$, and the covariance between the $i$-th and the $j$-th variable is $0.5^{|i-j|}$.
Denote the sample size by $n$, and the standard deviation of the error term by $\sigma$, we consider three scenarios: (1) $n=40$ and $\sigma=3$; (2) $n=40$ and $\sigma=1$; and (3) $n=60$ and $\sigma=1$. We repeat the simulation for 100 times and compare the results of Algorithm \ref{alg:componetwise}  and Algorithm \ref{alg:modified} with the results of LASSO, SCAD, and the Oracle model from \citet{Fan01VS}.

To evaluation the estimation accuracy, we compute the  model error (ME) by
\begin{equation} \label{def:ME}
 \text{ME} = (\hat{\bb}-\bb^{*})^{T}\Sigma(\hat{\bb}-\bb^{*})/\sigma^{2}
 \end{equation}
 where $\Sigma$ is the covariance matrix of the eight covariates. We set the ME from the ordinary least square (OLS) model as the benchmark, and compute the relative model errors: dividing ME of the other models by that of the OLS.  To obtain a robust criterion, we take the median of the relative model errors, namely the median of the relative model error (MRME). The results are shown in Table \ref{tab:eg1-MRME}. When $n$ gets larger or $\sigma$ gets smaller, the two VB algorithms and SCAD become much better in terms of MRME, while Lasso does not gain obvious improvement. Overall, the two VB algorithms have lower MRME than SCAD. When the sample size gets larger and the noise level gets smaller, the MRMEs of the two VB algorithms are approaching to that of the Oracle. Also, Algorithm \ref{alg:modified} is consistently better than Algorithm \ref{alg:componetwise}, especially when the sample size is small and the noise level is high.

To evaluation the selection accuracy, we count the number of zero coefficients among the signal and the noise variables. The results are also reported in Table \ref{tab:eg1-MRME}. The ``Correct''/``Incorrect'' column records the average number of zero-coefficients returned by the method among noise/signal variables. A good method should have a number close to $5$ for the ``Correct" column and $0$ for the ``Incorrect" column. First, we notice that there is a trade-off between ``Correct'' and ``Incorrect,'' or namely the trade-off between sensitivity and specificity of identifying  noise variables.  When the sample size is small ($n=40$) and the noise level is high ($\sigma=3$), Lasso identifies almost all the true predictors at the cost of including around 1.5 fake ones, while the variational Bayesian methods, on the other hand, correctly identify most noise variables at the cost of missing some signal variables. Hence, under such a low sample size and high variance setup, it is hard to tell which method is significantly better than the others.   But when the sample size gets larger and/or the noise gets smaller, the two VB algorithms outperform all the other methods except the Oracle.

\begin{table}[htbp]
\centering
\begin{tabular}{@{}cccc@{}}
\toprule
\multicolumn{1}{l}{}              & \multicolumn{1}{l}{}         & \multicolumn{2}{c}{\begin{tabular}[c]{@{}c@{}}Avg. No. of 0\\   Coefficients\end{tabular}} \\
\multicolumn{1}{l}{Method}        & \multicolumn{1}{l}{MRME(\%)} & \multicolumn{1}{l}{Correct}                 & \multicolumn{1}{l}{Incorrect}                \\ \midrule
\multicolumn{1}{l}{$n=40, \sigma=3$} &                              &                                             &                                              \\
SCAD                              & 72.90                        & 4.20                                        & 0.21                                         \\
Lasso                             & 63.19                        & 3.53                                        & 0.07                                         \\
Alg.1                             & 64.23                        & 4.36                                        & 0.32                                         \\
Alg.2                             & 60.27                        & 4.40                                        & 0.30                                         \\
Oracle                            & 33.31                        & 5                                           & 0                                            \\ \hline
\multicolumn{1}{l}{$n=40, \sigma=1$} &                              &                                             &                                              \\
SCAD                              & 54.81                        & 4.29                                        & 0                                            \\
Lasso                             & 63.19                        & 3.51                                        & 0                                            \\
Alg.1                             & 39.34                        & 4.85                                        & 0                                            \\
Alg.2                             & 37.37                        & 4.92                                        & 0.12                                            \\
Oracle                            & 33.31                        & 5                                           & 0                                            \\ \hline
\multicolumn{1}{l}{$n=60, \sigma=1$} &                              &                                             &                                              \\
SCAD                              & 47.54                        & 4.37                                        & 0                                            \\
Lasso                             & 65.22                        & 3.56                                        & 0                                            \\
Alg.1                             & 35.69                        & 4.92                                        & 0                                            \\
Alg.2                             & 34.74                        & 4.91                                        & 0                                            \\
Oracle                            & 29.82                        & 5                                           & 0                                            \\ \bottomrule

\end{tabular}
\caption{MRME and Average Number of 0 Coefficients. ``Correct'' represents how many noise variables are  correctly identified, ``Incorrect'' means the number of true predictors being erroneously set to zero. The results for LASSO, SCAD, and Oracle are  from \citet{Fan01VS}.}
\label{tab:eg1-MRME}
\end{table}

\subsection{Example 2: Highly Correlated Noisy Data}
This example is from \citet{wang11randomlasso}, in which the design matrix contains some highly correlated predictors with different signs. We use this example to demonstrate the advantage of the proposed  batch-wise update used by Algorithm \ref{alg:modified} over the component-wise update used by Algorithm \ref{alg:componetwise}: errors tend to accumulate with the component-wise update, especially when predictors are highly correlated.

The regression model has $p=40$ highly correlated covariates and the true coefficient vector is $\bb^{0}=(3, 3, -2, 3, 3, -2, 0,\ldots,0 )^{T} $ with the last $34$ elements being zero. The covariates are generated from a multivariate normal distribution: the variance of each variable is 1; the pairwise correlation of the first three variables is 0.9, that of the next three variables is 0.9, and all the other pairwise correlations are 0. The error terms are i.i.d. $N(0, 6^2)$. The sample size $n$ is either 50 or 100. For each $n$, we repeat the experiment 100 times and compute ME that is defined in (\ref{def:ME}). 

In Table \ref{tab:eg2-RME}, we report the average and the standard error of ME over 100 simulations. From the table, we can see that
Algorithm \ref{alg:modified} is better than Algorithm \ref{alg:componetwise}  for both $n=50$ and $n=100$. and it outperforms Lasso when the sample size grows.

\begin{table}[htbp]
\centering
\begin{tabular}{lcccc}
\toprule
      & OLS   & Lasso & Algorithm \ref{alg:componetwise}  & Algorithm \ref{alg:modified} \\ \midrule
$n=50$  & 4913  & 233   & 322         & 305         \\
      & (323) & (11)  & (17)        & (18)        \\ \cline{2-5}
$n=100$ & 706   & 144   & 139         & 109         \\
      & (25)  & (6)   & (7)         & (7)         \\ \bottomrule
\end{tabular}
\caption{$1000 \times$ Average ME of Different Methods. The numbers in the parentheses are the corresponding $1000\times$ standard errors. The results for LASSO are from  \citet{wang11randomlasso}. }
\label{tab:eg2-RME}
\end{table}

In Table \ref{tab:eg2-VSF}, we  list the minimum, the median, and the maximum of the selection frequencies in pairs of parentheses of the 6 important variables (IV) and the 34 unimportant variables (UV). Although  Algorithm \ref{alg:componetwise}  has the highest value for the maximum selection frequency of IV's, its median selection frequency of IV's is always the lowest. When the sample size increases, its median selection frequency of IV's even drops. Overall, Algorithm \ref{alg:modified} is better than Algorithm \ref{alg:componetwise}.
\begin{table}[htbp]
\centering
\begin{tabular}{@{}cccc@{}}
\toprule
\multicolumn{1}{l}{}      & Lasso        & Algorithm \ref{alg:componetwise}   & Algorithm \ref{alg:modified}  \\ \midrule
\multicolumn{1}{l}{$n=50$}  &              &              &              \\
IV                        & (11, 70, 77) & (24, 41 ,82) & (24,61,71)   \\
UV                        & (12, 17, 25) & (5, 11,17)  & (7,11,19)   \\ \hline
\multicolumn{1}{l}{$n=100$} &              &              &              \\
IV                        & (8, 84, 88)  & (12,  33, 99) & (18, 68, 73) \\
UV                        & (12, 22, 31) & (4,  7,  10)  & (0,  0,  2)  \\ \bottomrule
\end{tabular}
\caption{Variable Selection Frequencies (\%). IV: important variables; UV: unimportant variables. The three numbers in parentheses are the min, median, and max of selection frequencies among all important or unimportant variables, respectively. The results for LASSO are from  \citet{wang11randomlasso}.}
\label{tab:eg2-VSF}
\end{table}

In Table \ref{tab:eg2:coef-sign-50} and \ref{tab:eg2:coef-sign-100}, we compare the estimated coefficients, as well as their signs, from Lasso, Algorithm \ref{alg:componetwise}, and Algorithm \ref{alg:modified}. For $n=50$, the three methods can hardly detect any negative signs of $\beta_3$ or $\beta_6$: the high correlations between $\beta_2$ and $\beta_3$, and between $\beta_5$ and $\beta_6$ makes it difficult to identify the opposite signs of the neighboring highly-correlated predictors. When $n=100$, Algorithm \ref{alg:modified} performs the best. Firstly, it can identify the negative signs of $\beta_3$ and $\beta_6$ in some simulations, while Lasso and Algorithm \ref{alg:componetwise} can barely identify any negative signs of  $\beta_3$ and $\beta_6$. Secondly, the magnitude of the estimates from Algorithm \ref{alg:modified} is closer to the true coefficients, and the average coefficient estimates of $\beta_{3}$ and $\beta_{6}$ are negative, which is correct. Estimates from Lasso seem to deviate the most from the truth, which can also be confirmed by Table \ref{tab:eg2-RME}. 

\begin{table}[htbp]
\centering
\begin{tabular}{@{}ccccccc@{}}
\toprule
\multicolumn{1}{l}{}             & $\beta_1$   & $\beta_2$   & $\beta_3$   & $\beta_4$   & $\beta_5$   & $\beta_6$   \\ \midrule
\multicolumn{1}{l}{True Coef}    & 3      & 3      & -2     & 3      & 3      & -2     \\ \hline
\multicolumn{1}{l}{Lasso} &        &        &        &        &        &        \\
Ave. of est.                     & 1.41   & 1.27   & 0.12   & 1.36   & 1.36   & 0.09   \\
                                 & (0.12) & (0.14) & (0.04) & (0.12) & (0.13) & (0.06) \\
No. of pos. sgn.                 & 73     & 69     & 13     & 74     & 68     & 13     \\
No. of neg. sgn.                 & 0      & 0      & 1      & 0      & 0      & 2      \\ \hline
\multicolumn{1}{l}{Alg.1} &        &        &        &        &        &        \\
Ave. of est.                     & 2.08   & 0.75   & 0.23   & 2.19   & 0.61   & 0.23   \\
                                 & (0.16) & (0.11) & (0.05) & (0.16) & (0.09) & (0.04) \\
No. of pos. sgn.                 & 80     & 43     & 24     & 82     & 40     & 26     \\
No. of neg. sgn.                 & 0      & 0      & 0      & 0      & 0      & 0      \\ \hline
\multicolumn{1}{l}{Alg.2} &        &        &        &        &        &        \\
Ave. of est.                     & 1.63   & 1.17   & 0.30   & 1.36   & 1.44   & 0.28   \\
                                 & (0.16) & (0.13) & (0.09) & (0.14) & (0.14) & (0.07) \\
No. of pos. sgn.                 & 71     & 57     & 31     & 65     & 66     & 24     \\
No. of neg. sgn.                 & 0      & 0      & 1      & 0      & 0      & 0      \\ \bottomrule
\end{tabular}
\caption{Coefficient and Coefficient Sign Estimation, $n=50$. The numbers in parenthesis are the standard errors. The results for LASSO are from  \citet{wang11randomlasso}.}
\label{tab:eg2:coef-sign-50}
\end{table}

\begin{table}[htbp]
\centering
\begin{tabular}{@{}ccccccc@{}}
\toprule
\multicolumn{1}{l}{}              & $\beta_1$   & $\beta_2$   & $\beta_3$   & $\beta_4$   & $\beta_5$   & $\beta_6$   \\ \midrule
\multicolumn{1}{l}{True Coef}     & 3      & 3      & -2     & 3      & 3      & -2     \\ \hline
\multicolumn{1}{l}{Lasso} &        &        &        &        &        &        \\
Ave. of est.                      & 1.67   & 1.50   & 0.06   & 1.85   & 1.38   & 0.04   \\
                                  & (0.11) & (0.11) & (0.02) & (0.13) & (0.13) & (0.03) \\
No. of pos. sgn.                  & 91     & 85     & 8      & 86     & 78     & 9      \\
No. of neg. sgn.                  & 0      & 0      & 0      & 0      & 0      & 2      \\ \hline
\multicolumn{1}{l}{Alg.1} &        &        &        &        &        &        \\
Ave. of est.                      &2.97 & 0.56 & 0.08 & 2.93 & 0.56 & 0.10 \\
                                  & (0.12) & (0.08) & (0.02) & (0.12) & (0.09) & (0.03) \\
No. of pos. sgn.                  & 99     & 34     & 12      & 96     & 32     & 12      \\
No. of neg. sgn.                  & 0      & 0      & 0      & 0      & 0      & 0      \\ \hline
\multicolumn{1}{l}{Alg.2} &        &        &        &        &        &        \\
Ave. of est.                      & 2.09   & 1.92   & -0.11  & 2.33   & 1.85   & -0.20  \\
                                  & (0.16) & (0.16) & (0.05) & (0.18) & (0.18) & (0.09) \\
No. of pos. sgn.                  & 73     & 66     & 5      & 72     & 65     & 7     \\
No. of neg. sgn.                  & 0      & 1      & 13      & 0      & 3      & 16      \\ \bottomrule
\end{tabular}
\caption{Coefficient and Coefficient Sign Estimation, $n=100$. The numbers in parenthesis are the standard errors. The results for LASSO are from  \citet{wang11randomlasso}.}
\label{tab:eg2:coef-sign-100}

\end{table}

\subsection{Example 3: Large $p$, Small $n$}
We first consider a large-$p$-small-$n$ example from \citet{George14EMVS},
 in which only the first three of $p=1000$ predictors are the true ones with non-zero coefficients to be 3, 2, and 1 respectively, and the sample size $n=100.$ The covariance between the $i$-th and $j$-th variables is $0.6^{|i-j|}$, and the error terms are generated from $i.i.d.$ $N(0, 3)$.

\begin{figure}[htbp]
\centering
\begin{subfigure}{.5\textwidth}
  \centering
   \includegraphics[width=\linewidth, page = 1]{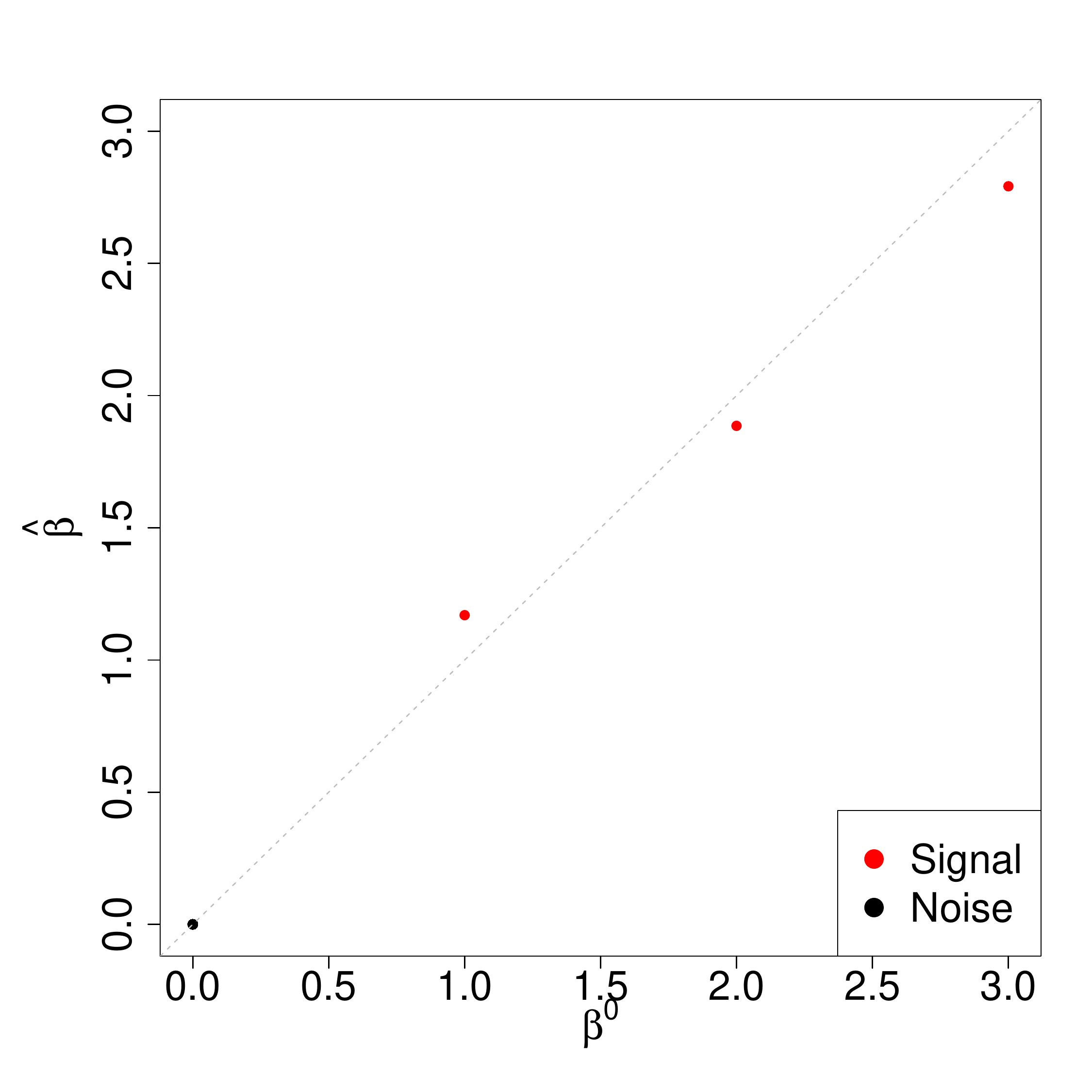}
  \caption{True Coefficient vs. Sparse Estimates: Algorithm \ref{alg:componetwise} }
  \label{fig:sim3-Geroge-alg1}
\end{subfigure}
\begin{subfigure}{.5\textwidth}
  \centering
  \includegraphics[width=\linewidth, page = 2]{EMVS_posterior_mode_vs_true.pdf}
  \caption{True Coefficient vs. Sparse Estimates: Algorithm \ref{alg:modified}}
  \label{fig:sim3-Geroge-alg2}
\end{subfigure}
\caption{$\hat{\bb}$ \textit{vs.} $\bb^{0}$ for the large-$p$-small-$n$ example from \citet{George14EMVS}.}
\end{figure}

We fit the model using Algorithm \ref{alg:componetwise}  and Algorithm \ref{alg:modified}, with fixed $v_1=1$. As shown in Figure \ref{fig:sim3-Geroge-alg1} and \ref{fig:sim3-Geroge-alg2}, both algorithms make no mistake in variable selection and the estimates are very close to the true coefficients.

Next we make this example a little more challenging by adding more non-zero coefficients of different magnitudes. The true coefficient vector is now set to be $\bb^{0}=({\bb^{0}_{1}}^{T}, 0, \ldots,0)^{T},$ where the last 980 elements are all zero and the 20 non-zero coefficients in $\bb^{0}_{1}$ contains randomly distributed ten 1's, seven 2's, and three 3's.
With the 0.6 pairwise correlation, some weak signals may be overshadowed by nearby strong signals, which makes variable selection a challenging task. We repeat the experiment 100 times, and compare results from  Algorithm \ref{alg:componetwise}, Algorithm \ref{alg:modified} and Lasso. For the tuning parameter in Lasso, we use the default \texttt{cv.glmnet} function from the \texttt{R} package \texttt{glmnet} \citep{friedman10glmnet}, and report the results for both \texttt{lambda.min} that is the $\lambda$ with the smallest CV error, and \texttt{lambda.1se} that is the largest $\lambda$ whose CV error is within one standard error of the smallest CV error.

The results are summarized in Table \ref{tab:sim3}. Regarding the accuracy for identifying important variables, Algorithm \ref{alg:componetwise}  is the worst: on average it only identifies 8.97 true predictors, fewer than a half. This is supported with our large sample analysis: with high correlation, the errors tend to accumulate, which makes some true predictors hard to be identified. Regarding the accuracy for identifying unimportant variables, \texttt{Lasso.min} is the worst: on average it selects 34.14 noise variables, more than doubled that of \texttt{Lasso.1se}, the second worst. Overall, Algorithm \ref{alg:modified} performs the best.

\begin{table}[htbp]
\centering
\begin{tabular}{@{}clcccc@{}}
\toprule
\multicolumn{1}{l}{}      & Model & \multicolumn{1}{l}{Algorithm \ref{alg:componetwise} } & \multicolumn{1}{l}{Algorithm \ref{alg:modified}} & \multicolumn{1}{l}{Lasso.min} & \multicolumn{1}{l}{Lasso.1se} \\ \cmidrule(l){2-6}
\multirow{2}{*}{\# of IV} & Mean  & 8.97                            & 15.61                           & 19.92                         & 19.77                         \\
                          & S.E.  & (0.13)                          & (0.19)                          & (0.03)                        & (0.05)                        \\ \cmidrule(l){2-6}
\multirow{2}{*}{\# of UV} & Mean  & 0.12                            & 0.5                             & 34.14                         & 11.23                         \\
                          & S.E.  & (0.04)                          & (0.08)                          & (1.50)                        & (0.81)                        \\ \bottomrule
\end{tabular}
\caption{Variable Selection Frequencies (\%). IV: important variables; UV: unimportant variables.
The numbers in parenthesis are the standard errors. }
\label{tab:sim3}
\end{table}

\section{Real Data: Boston Housing Data}
\subsection{Introduction}
The original data is from the \texttt{R} library \texttt{mlbench}, which has 506 observations on 19 variables. We apply some suggested transformations on the data according to \citet{hardle07appliedmultivariate}, then remove three variables \texttt{medv}, \texttt{town}, and \texttt{tract}, and use \texttt{cmedv} as the response variable. We call this data set ``Boston Housing 1."

Then we create a larger data set called ``Boston Housing 2.'' First 
we add all the 119 quadratic terms (including all pairwise interaction terms) of the predictors, and 500 noise features. We generate the noise features in 50 batches. For each batch, we randomly select 10 variables from the set of 134 ``true'' variables, which gives us a $506 \times 10$ data matrix; for each entry of the matrix, we add a small Gaussian error, and then randomly shuffle the $506$ rows. So each noise feature looks like some true variable marginally, and in addition  correlations among the true variables are preserved in the noise features. The final data set has 634 features, which is larger than the sample size.

As we have shown that Algorithm \ref{alg:modified} is better than Algorithm \ref{alg:componetwise}  in the simulation study, we just compare Algorithm \ref{alg:modified} to Lasso, Ridge, and the full model (i.e., the model using all predictors). For Algorithm \ref{alg:modified}, we consider the following two prediction methods:
\begin{enumerate}[(i)]
\item Use the sparse estimate $\hat{\bb}$ that is defined at (\ref{eq:sparse:beta}) and the prediction is given by $\hat{\y} = \X^{new} \hat{\bb}$.
We abbreviate this approach by ``S'' that stands for  ``Sparsity Prediction.''
  \item Refit a linear regression model using the predictors from $\hat{S} = \{j: \phi_{j}>0.5\}$ and  denote the estimated coefficients from the OLS by $\hat{\bb}^{ols}$. Then the prediction is given by $\hat{\y} = \X^{new}_{[,\hat{S}]} \hat{\bb}^{ols}$, where  $\X^{new}_{[,\hat{S}]}$ denotes a subset of the data matrix $\X^{new}$ with only columns from $\hat{S}.$ We abbreviate this approach by ``TS'' that stands for ``Two-Stage Sparse Prediction''.
\end{enumerate}

We run the following simulation for 50 times. In each iteration, we randomly subset $75\%$ of the dataset as the training data (380 observations) and  predict on the remaining validation data (126 observations). Then we record the selected model size and compute the mean squared prediction error (MSPE). As all the methods but the OLS have a tuning parameter, we select the tuning parameter using cross-validation. For Algorithm \ref{alg:modified}, we use a 5-fold cross validation for the two prediction approachs;  for Lasso and ridge regression, we use the default \texttt{cv.glmnet} function from the \texttt{R} package \texttt{glmnet} \citep{friedman10glmnet}, and report results for both \texttt{lambda.min}
and \texttt{lambda.1se}.
The results are summarized in Table \ref{tab:bh}.

\subsection{Boston Housing 1}
Most methods perform similarly based on the prediction error. Surprisingly the full model performs the best. This is because the potential gain of variable selection for such a traditional small-$p$-large-$n$ example is negligible. On the other hand, the potential bias introduced by variable selection or shrinkage procedures, when relevant variables are mistakenly excluded or over-shrunk, can be large. This also explains why Lasso.1se/ridge.1se performs worse than Lasso.min/ridge.min, since the former tends to pick a smaller model than the latter, i.e., has a higher chance of missing relevant variables.


\subsection{Boston Housing 2}
The ridge regression is the worst of all methods: both ridge.1se and ridge.min have relatively large effective dimensions, but high prediction errors. 
For prediction accuracy, Lasso.min, Alg2.S and Alg2.TS are better than the other methods; regarding sparsity, the models selected by Lasso.1se, Alg2.S and Alg2.TS are much smaller than the others. The best model is Alg2.TS: it has the best prediction accuracy with the most sparse model. 

\begin{table}[htbp]
  \centering
     \begin{tabular}{cccccccccc}
    \toprule
          & Model & Full & Ridge.min & Ridge.1se & Lasso.min & Lasso.1se  & Alg2.S  & Alg2.TS \\
    \midrule
    BH 1 & Mean & 0.043 & 0.044 & 0.049 & 0.044 & 0.048 & 0.045 & 0.044 \\
          & SE & (0.008)  & (0.009)  & (0.010)  & (0.008)  & (0.008)  & (0.009)    & (0.008)  \\
          & Size & 15  & 12.34 & 9.21  & 13.84 & 7.2    & 10.8      & 11.36 \\
	\hline \\
    BH 2 & Mean &       & 0.065 & 0.071 & 0.043 & 0.047 & 0.046  & 0.042 \\
          & SE &       & (0.011) & (0.012) &  (0.008)  & (0.008)  & (0.012)    & (0.007)  \\
          & Size &       & 52.21 & 40.21   & 38.7  & 8.68  & 9.18   & 7.74 \\
    \bottomrule
\end{tabular}
 \caption{Boston Housing Data: Average and S.E. of MSPE, and Model Size}
  \label{tab:bh}%
\end{table}%


\section{Conclusions}
The Bayesian approach to variable selection is appealing since it outputs not only a single model but a probability distribution over all possible models. Hence model uncertainty can be naturally incorporated into estimation, prediction, and many other statistical inferences. However, most Bayesian variable selection methods are implemented through MCMC, which is time consuming when the model dimension is large. In this paper, we propose an algorithm that  approximates the posterior distribution via a variational optimization. Our proposed algorithm converges very fast and can scale up with large data sets. We also showed that the approximation returned by our algorithm has the desired asymptotic behavior, which achieves both the frequentist consistency and Bayesian consistency asymptotically.

\newpage
\appendix

\section*{Appendices}

\subsection*{Appendix A. Update $\phi_j$'s in Algorithm \ref{alg:modified}}

We provide the detailed derivation of the updating equation for $\phi_{j}$'s here. The derivation of other variational distributions and MAP estimators is straightforward. 

We fix $\{\mu_{j}\}_{j=1}^{p}$, $\{\sigma_{j}^{2}\}_{j=1}^{p}$, $\hat{\theta}$, and $\hat{\sigma}^{2}$, and update $\{\phi_{j}\}_{j=1}^{p}$. 
The objective function is given by
\begin{align*}
    \Omega(\phi_{1},\ldots, \phi_{p}) 
    =&  -\frac{1}{2\hat{\sigma}^{2}}\left[ \y^{T}\y - 2\y^{T}\X\E(\bb) + \E\left(\bb^{T}(\X^{T}\X)\bb\right) \right] \\
    & + \sum_{j=1}^{p}\phi_{j}\left( \frac{1}{2}\log\frac{\sigma_{j}^{2}}{v_{1}\hat{\sigma}^{2}} + \frac{1}{2} -\frac{\mu_{j}^{2} + \sigma_{j}^{2}}{2v_{1}\hat{\sigma}^{2}} + \log(\hat{\theta}) -  \log(\phi_{j})  \right)\\
    &\quad +  (1-\phi_{j})[\log(1-\hat{\theta}) - \log(1-\phi_{j})] + \mathrm{Constant}.
\end{align*}
Denote $\U = \diag\{\mu_{1},\ldots, \mu_{p}\}$ and $\pp = (\phi_{1}, \ldots, \phi_{p})^{T}$, we have $\E(\bb) = \U\pp$ and
\begin{align*}
& \E\left(\bb^{T}(\X^{T}\X)\bb\right) \\
=& \tr\left((\X^{T}\X) \Cov(\bb)\right) + \E(\bb) ^{T}(\X^{T}\X)\E(\bb) \\
=& \sum_{j=1}^{p} n (\mu_{j}^{2} + \sigma_{j}^{2}) \phi_{j} - n \pp^{T} \U \U\pp +\pp^{T} \U(\X^{T}\X) \U\pp .
\end{align*}
Then the objective function becomes
\begin{align*}
\Omega(\phi_{1},\ldots, \phi_{p}) = &  \frac{1}{\hat{\sigma}^{2}}\y^{T}\X\U\pp - \frac{n}{2\hat{\sigma}^{2}}\sum_{j=1}^{p}(\mu_{j}^{2} + \sigma_{j}^{2}) \phi_{j} \\
&- \frac{1}{2\hat{\sigma}^{2}} \pp^{T} \U(\X^{T}\X - n \mathbf{I} ) \U\pp \\
& + \sum_{j=1}^{p}\phi_{j}\left( \frac{1}{2}\log\frac{\sigma_{j}^{2}}{v_{1}\hat{\sigma}^{2}} + \frac{1}{2} -\frac{\mu_{j}^{2} + \sigma_{j}^{2}}{2v_{1}\hat{\sigma}^{2}} + \log(\hat{\theta}) -  \log(\phi_{j})  \right)\\
    &\quad +  (1-\phi_{j})[\log(1-\hat{\theta}) - \log(1-\phi_{j})]  + \mathrm{Constant}.
\end{align*}
Taking derivative w.r.t. $\pp$, we have
\begin{align*}
 \nabla\Omega = &  \frac{1}{\hat{\sigma}^{2}}\y^{T}\X\U - \frac{n}{2\hat{\sigma}^{2}}\{(\mu_{j}^{2} + \sigma_{j}^{2}) \}_{j=1}^{p} \\
&- \frac{1}{\hat{\sigma}^{2}}  \U(\X^{T}\X - n \mathbf{I} ) \U\pp + Rest.
\end{align*}
Direct optimization would involve numerical methods due to the non-linear system with constraints. Thus, we take an approximation approach.
Denote $\Delta = \U(\X^{T}\X - n\I_{p}\}) \U$ and define $g(\pp) = \pp^{T}\Delta\pp$. At the $t$-th iteration, using the Taylor expansion, we approximate the quadratic form $g(\pp^{(t)})$ by
\begin{align*}
g(\pp^{(t)}) \approx& g(\pp^{(t-1)}) + \nabla(g(\pp^{(t-1)}))^{T}(\pp^{(t)} - \pp^{(t-1)}) \\
=& (\pp^{(t-1)})^{T} \Delta \pp^{(t-1)} + 2(\pp^{(t-1)})^{T}\Delta(\pp^{(t)} - \pp^{(t-1)}) \\
\propto& 2(\pp^{(t-1)})^{T}\Delta \pp^{(t)}.
\end{align*}
Hence, we have
\begin{align*}
\Omega \approx &   \frac{1}{\hat{\sigma}^{2}}\y^{T}\X\U\pp - \frac{n}{2\hat{\sigma}^{2}}\sum_{j=1}^{p}(\mu_{j}^{2} + \sigma_{j}^{2})\phi_{j} \\
& - \frac{1}{\hat{\sigma}^{2}} (\pp^{(t-1)})^{T}\U(\X^{T}\X - n \mathbf{I} )\U \pp \\
& + \sum_{j=1}^{p}\phi_{j}\left( \frac{1}{2}\log\frac{\sigma_{j}^{2}}{v_{1}\hat{\sigma}^{2}} + \frac{1}{2} -\frac{\mu_{j}^{2} + \sigma_{j}^{2}}{2v_{1}\hat{\sigma}^{2}} + \log(\hat{\theta}) -  \log(\phi_{j})  \right)\\
    &\quad +  (1-\phi_{j})[\log(1-\hat{\theta}) - \log(1-\phi_{j})] + \mathrm{Constant}.
\end{align*}
Taking partial derivative w.r.t.\! $\phi_{j}$'s respectively, we have
\begin{align*}
\frac{\partial \Omega}{\partial\phi_{j}} \approx &  \frac{1}{\hat{\sigma}^{2}}\y^{T}\X_{j}\mu_{j} - \frac{n}{2\hat{\sigma}^{2}}(\mu_{j}^{2} + \sigma_{j}^{2}) 
- \frac{1}{\hat{\sigma}^{2}} \sum_{k\neq j} \mu_{j}\mu_{k}\X_{k}^{T}\X_{j}\phi_{k}^{(t-1)} \\
& + \frac{1}{2}\log\frac{\sigma_{j}^{2}}{v_{1}\hat{\sigma}^{2}} + \frac{1}{2} -\frac{\mu_{j}^{2} + \sigma_{j}^{2}}{2v_{1}\hat{\sigma}^{2}} + \log(\hat{\theta}) -  \log(\phi_{j}) -1
 - \log(1-\hat{\theta}) + \log(1-\phi_{j}) +1. 
 \end{align*}
 Setting $\frac{\partial \Omega}{\partial\phi_{j}} =0$, we have
\begin{equation} 
\Logit(\phi_{j}) =  \Logit(\hat{\theta}) + \frac{1}{2}\log\frac{\sigma_{j}^{2}}{v_{1}\hat{\sigma}^{2}} - \frac{\mu_{j}^{2}}{2\sigma_{j}^{2}} + \frac{1}{\hat{\sigma}^{2}}\mu_{j}\X_{j}^{T}\left(\y-\X_{[,-j]}\PP^{(t-1)}_{[-j,-j]}\uu_{[-j]}\right), \label{eq:alg2-phi-update-temp}
\end{equation}
where the last term is equal to $\frac{1}{\hat{\sigma}^{2}}\mu_{j}\X_{j}^{T} \mu_{j}(n +1/v_{1})$ according to Algorithm \ref{alg:componetwise}. Therefore, we further approximate it by $\frac{1}{\hat{\sigma}^{2}}\mu_{j}^{2}(n+1/v_{1})$ to reduce computational complexity. Then we have
\begin{align*}
\Logit(\phi_{j}) =& \Logit(\hat{\theta}) + \frac{1}{2}\log\frac{\sigma_{j}^{2}}{v_{1}\hat{\sigma}^{2}} + \frac{\mu_{j}^{2}}{2\hat{\sigma}^{2}}\left(n +\frac{1}{v_{1}}\right) \\
=& \Logit(\hat{\theta}) + \frac{1}{2}\log\frac{\sigma_{j}^{2}}{v_{1}\hat{\sigma}^{2}} + \frac{\mu_{j}^{2}}{2\sigma_{j}^{2}}.
\end{align*}

\subsection*{Appendix B. Proofs for Section 3}
\subsubsection*{B.1 Proof for Lemma \ref{lemma:one-step-gap}}
In Algorithm \ref{alg:modified}, we first update $\{\mu_{nj}\}_{j=1}^{p}$, given the initial value $\phi_{j}=1$. The updating formula for $\{\mu_{nj}\}_{j=1}^{p}$ at the first iteration is
\begin{align*}
  \uu_{n} =& \left(\PP\X^{T}\X\PP + n\PP(\I_{p}-\PP) + \frac{1}{v_{1}} \PP\right)^{-1} \PP\X^{T}\y_{n} \\
  =& \left[\X^{T}\X + \frac{1}{v_{1}} \I_{p}\right]^{-1} \X^{T}(\X\bb^{*} + \ee_{n}),
\end{align*}
where $\PP=\I_{p}$. After updating $\uu_{n}$, given the initial values $\hat{\theta}=1/2$ and $\hat{\sigma}^2=1$, we update the logit of $\phi_{j}$ using
\begin{align*}
2\Logit(\phi_{j})
= - \log(a_n v_{1} + 1) + \mu_{nj}^{2} \left(a_n+\frac{1}{v_{1}}\right).
\end{align*}

To quantify the magnitude of $\Logit(\phi_{j})$, the key is to quantify $\mu_{nj}$. Decompose $\uu_{n}$ into three parts: the true coefficient vector $\bb^{*}$, the bias $\b_{n}$, and the error projection $\w_{n}$.
\begin{align*}
\uu_{n} =& \bb^{*} - \left(v_1 \X^{T}\X +  \I_{p}\right)^{-1} \bb^{*} + \left(\X^{T}\X +  \frac{1}{v_1} \I_{p} \right)^{-1} \X^{T} \ee_{n} \\
=& \bb^{*} - \b_{n} + \w_{n}.
\end{align*}
Next we prove the following results.
\begin{enumerate}
  \item \textbf{Bound for $\b_{n}$.}
Denote the singular value decomposition of $\X$ as $PDQ^T$, where $r$ is the rank of $\X$, the dimension of $P$, $D$, and $Q$ are $n\times r$, $r\times r$, and $p\times r$, respectively. Condition (C1) implies that the true coefficient vector $\bb^*$ is the projection of the set $\mathcal{B}$ onto the row space of $\X$, i.e., $\bb^*=QQ^T\bb^*$. Then the bias term can be written as
\begin{eqnarray*}
\b_n &=&\big (v_1 \X^T \X + \I_p  \big )^{-1} \bb^* \\
&=& (v_1 QD^2Q^T+\I_p)^{-1}QQ^T\bb^*\\
&=& Q(v_1 D^2+\I_r)^{-1}Q^T\bb^*.
\end{eqnarray*}
So we can bound the maximal of the bias term $b_{nj}$ by
    \begin{eqnarray}
      \max_{j}b_{nj}^{2} &\leq & \|\b_{n}\|_{2}^{2} = \|Q(v_1 D^2+\I_r)^{-1}Q^T\bb^*\|_{2}^{2} \nonumber \\
     & \leq &  \left(\frac{1}{v_1 \lambda_{n1} + 1 } \right)^{2} \|\bb^{*}\|_{2}^{2} = \frac{O(n^{\eta_{2}})}{(v_1 \lambda_{n1})^2}.  \label{eq:bound:b_nj}
    \end{eqnarray}

  \item \textbf{Bound for the variance of $w_{nj}$.} $\w_{n}$ is a Gaussian random variable with mean zero and covariance
    \begin{align*}
        \Cov(\w_{n}) =& \Cov((\X^{T}\X +  \frac{1}{v_1} \I_{p})^{-1} \X^{T} \ee_{n}) \\
     =& \sigma^2 Q\diag\left(\frac{d_j^2}{(d_j^2+1/v_1)^2}\right)Q^T
    \end{align*}
    where $d_j$'s are the elements from the diagonal matrix $D$.  So the variance of each $w_{nj}$ is bounded by the largest eigenvalue of $\Cov(\w_{n})$:
        \begin{equation}
       \text{Var}(w_{nj}) \le  \frac{\sigma^{2}\lambda_{n1}}{(\lambda_{n1}+1/v_1)^2} \leq \frac{\sigma^{2}}{\lambda_{n1}}.
    \end{equation}

\item \textbf{Inequalities for $\mu_{nj}^2.$} Using the inequality $(\sum_{i=1}^{n}a_{i})^{2} \leq n\sum_{i=1}^{n}a_{i}^{2}$, we have
\begin{eqnarray*}
\max_{j \notin S^*_n} \mu_{nj}^{2} &\leq &  2(\max_{j}b_{nj}^{2} + \max_{j}w_{nj}^{2}), \\
\min_{j\in S_{n}^{*}} \mu_{nj}^{2} & \geq & \frac{1}{3}\min_{j\in S_{n}^{*}}(\beta_{j}^{*})^{2} - \max_{j}b_{nj}^{2} - \max_{j}w_{nj}^{2}.
\end{eqnarray*}
\end{enumerate}

First we  show (\ref{lemma:prob:noise}). Note
 \begin{equation*}
      \max_{j\notin S_{n}^{*}} 2\Logit(\phi_{j}) \leq -\log(a_n v_{1} +  1) + 2 \Big ( a_n + \frac{1}{v_1} \Big) \Big ( \max_{j}b_{nj}^{2} + \max_{j}w_{nj}^{2} \Big).
\end{equation*}
We have
\begin{enumerate}[(a)]
   \item $\log(v_{1} a_n + 1) \to \infty$ since $v_1 \succ n^{-a}.$
      \item By (\ref{eq:bound:b_nj}), $( a_n + \frac{1}{v_1}) \max_{j}b_{nj}^{2} = O(n^{\eta_2+a-2 \eta_1}/v_1^2 )\to 0$ since $v_1 \succ n^{- \frac{1}{2}(2 \eta_1 - a-\eta_2)}.$
   \item  For $(a_n + 1/v_1) w_{nj}^{2}$, its variance is upper bounded by $O(n^{a-\eta_1})$. Hence, for any constant $c$,
   \begin{align*}
 & P\left( (a_n + 1/v_1) \max_{j}w_{nj}^{2} > c \right) \\
\leq & \sum_{j=1}^p  P\left ( (a_n + 1/v_1) w_{nj}^{2} > c \right) \\
\leq & \frac{p}{\sqrt{2 \pi n^{\eta_1 - a}}} \exp\left\{-\frac{c}{2} n^{\eta_1 - a}  \right \} \longrightarrow 0,
\end{align*}
since $\log p = o(n^{\eta_1-a}). $ 
\end{enumerate}
Thus, $\max_{j\notin S_{n}^{*}} 2\Logit(\phi_{j}) \overset{P}{\to} -\infty$ as $n \to \infty$, and therefore (\ref{lemma:prob:noise}) holds true. 

Next we show (\ref{lemma:prob:signal}). Note
\[
\min_{j\in S_{n}^{*}}2\Logit(\phi_{nj}) \geq -\log(a_n v_{1} +  1) + \Big ( a_n + \frac{1}{v_1} \Big) \Big ( \frac{1}{3}\min_{j\in S_{n}^{*}}(\beta_{j}^{*})^{2} - \max_{j}b_{nj}^{2} - \max_{j}w_{nj}^{2} \Big).
\]
Since $\left(a_n + \frac{1}{v_1} \right)\frac{1}{3}\min_{j\in S_{n}^{*}}(\beta_{j}^{*})^{2} \succeq n^{a-(1-\eta_3)} \succ \log(a_n v_1 + 1)$, it is the leading term. So  $\min_{j\in S_{n}^{*}}2\Logit(\phi_{nj}) \overset{P}{\to} \infty$ as $n \to \infty$ and therefore (\ref{lemma:prob:signal}) holds true.

\subsubsection*{B.2 Proof for Theorem \ref{thm:one-step-frequentist}}
With Lemma \ref{lemma:one-step-gap}, it is easy to show that when the sample size is large enough, our algorithm will stop with one update. For any threshold value $c$ in (\ref{eq:alg2-phi-update}), we can set $C$ in Lemma  \ref{lemma:one-step-gap} to be bigger than $2 \log (\frac{1}{c}-1)$. Then with probability going to $1$, after the first iteration, we will have $\max_{j\notin S_{n}^{*}} \phi_{j} < c$, and  $\min_{j\in S_{n}^{*}}\phi_{nj} > 1 - c$ and the algorithm will halt. Therefore, $\hat{S}_n = S_n^*$ with probability $1$ when $n \to \infty$. Hence, the frequentist selection consistency follows.

\subsubsection*{B.3 Proof for Theorem \ref{thm:one-step-bayesian}}
Let $C_{n}=s\log(v_{1}a_n)$, where $s \in (0,1)$. Following the same argument used in the proof of Lemma \ref{lemma:one-step-gap}, we can show that
\[
P\left( 2\max_{j\notin S_{n}^{*}}\Logit(\phi_{j}) > -C_{n}\right) \to 0 \quad \mathrm{and} \quad
P\left( 2\min_{j\in S_{n}^{*}}\Logit(\phi_{j}) < C_{n}\right) \to 0,
\]
if $\log p = o(n^{\eta_1 - a}).$
Therefore, with probability going to 1, we have
\begin{align*}
\max_{j\notin S_{n}^{*}}\log\frac{\phi_{j}}{1-\phi_{j}} < -\frac{1}{2}C_{n}  \Rightarrow& \max_{j\notin S_{n}^{*}}\phi_{j} <  \frac{1}{\exp\{\frac{1}{2}C_{n}\}}, \\
\min_{j\in S_{n}^{*}}\log\frac{\phi_{j}}{1-\phi_{j}} > \frac{1}{2}C_{n}  \Rightarrow&  \max_{j\in S_{n}^{*}}(1-\phi_{j})  < \frac{1}{\exp\{\frac{1}{2}C_{n}\}}.
\end{align*}
Using the inequality $\prod_{j}(1-p_{j}) \geq 1-\sum_{j} p_{j}$, we have
\begin{equation*}
1-q(\rr^*)
\leq  \sum_{j\in S_{n}^{*}}(1-\phi_{j}) + \sum_{j\notin S_{n}^{*}} \phi_{j}
\leq p \times \left[\max_{j\in S_{n}^{*}}(1-\phi_{j}) \bigvee \max_{j\notin S_{n}^{*}} \phi_{j} \right]
= \frac{p}{(v_{1}a_n)^{s/2}}.
\end{equation*}
If $p \prec (v_{1}a_n)^{s/2}$, we have
\begin{equation}
1-q(\rr^*) \leq  \frac{p}{e^{\frac{1}{2}C_{n}}} = \frac{p}{(v_{1}a_n)^{s/2}} \overset{P}{\longrightarrow} 0,
\end{equation}
as $v_1 \prec \exp(n^{a-(1-\eta_3)})$, $p \prec (v_1 a_n)^{s/2} \prec \exp(\sqrt{n}^{\frac{a -1+ \eta_3}{2}}) n^{s/2}$. This implies that if $p = o(\exp(n^{\frac{a-1-\eta_3}{2}}))$, we can achieve Bayesian consistency by letting $v_1$ going to infinity at an exponential order.

\newpage
\bibliography{Reference}

\end{document}